%% file: Gain-Belief-Sufficient-Condition.tex
\documentclass[submission,copyright,creativecommons]{eptcs}
\providecommand{\event}{TARK 2023} 

\usepackage{iftex}

\usepackage[table, dvipsnames]{xcolor}
\usepackage{enumerate}

\usepackage{pgf, tikz}
\usetikzlibrary{arrows, automata}

\usepackage{epigraph}

\usepackage{colonequals}
\usepackage{amssymb}
\usepackage{amsmath}
\usepackage{paralist}
\usepackage{amsthm}
\usepackage{adjustbox}

\usepackage{textcomp}

\usepackage{algorithm}
\usepackage[noend]{algpseudocode}

\usepackage{mathrsfs}
\usepackage{tikz}
\usepackage{tikz-cd}
\usetikzlibrary{shapes}
\usetikzlibrary{trees}
\usepackage{forest}
\usetikzlibrary{positioning,arrows,calc,matrix,shapes.multipart}
\tikzset{
modal/.style={>=stealth’,shorten >=1pt,shorten <=1pt,auto,node distance=1.5cm,
semithick},
world/.style={circle,draw,minimum size=0.5cm,fill=gray!15},
point/.style={circle,draw,inner sep=0.5mm,fill=black},
reflexive above/.style={->,loop,looseness=7,in=120,out=60},
reflexive below/.style={->,loop,looseness=7,in=240,out=300},
reflexive left/.style={->,loop,looseness=7,in=150,out=210},
reflexive right/.style={->,loop,looseness=7,in=30,out=330}
}

\usepackage{hyperref}
\usepackage[capitalise]{cleveref}

\usepackage{tabularx}

\usepackage[shortlabels]{enumitem}

\newtheorem{theorem}{Theorem}
\newtheorem{lemma}[theorem]{Lemma}
\newtheorem{corollary}[theorem]{Corollary}
\theoremstyle{definition} 
\newtheorem{definition}[theorem]{Definition}
\theoremstyle{remark}

\usepackage{stmaryrd}

\usepackage{footnotehyper}

\usepackage[appendix=append]{apxproof}
\usepackage[append]{optional}

\newtheoremrep{theorem}{Theorem}[section]
\newtheoremrep{lemma}[theorem]{Lemma}
\newtheoremrep{corollary}[theorem]{Corollary}

\newcommand{\bbbt}{\mathbb{T}}
\newcommand{\bbbn}{\mathbb{N}}
\newcommand{\ce}{\colonequals}
\newcommand{\cce}{\coloncolonequals}

\newcommand{\extension}[1][]{\mathscr{E}^{#1}}

\newcommand{\gevents}[1][]{\mathit{GEvents}_{#1}}
\newcommand{\bevents}[1][]{\mathit{BEvents}_{#1}}
\newcommand{\fevents}[1][]{\mathit{FEvents}_{#1}}
\newcommand{\sevents}[1][]{\mathit{SysEvents}_{#1}}
\newcommand{\gtrueevents}[1][]{\overline{\mathit{GEvents}}_{#1}}
\newcommand{\gtrueactions}[1][]{\overline{\mathit{GActions}}_{#1}}
\newcommand{\gactions}[1][]{\gtrueactions[#1]}

\newcommand{\gtruethings}[1][]{\overline{\ghaps[#1]}}

\newcommand{\events}[1][]{{\mathit{Events}}_{#1}}
\newcommand{\actions}[1][]{{\mathit{Actions}}_{#1}}
\newcommand{\haps}[1][]{{\mathit{Haps}}_{#1}}
\newcommand{\ghaps}[1][]{{\mathit{GHaps}}_{#1}}

\newcommand{\agents}{\mathcal{A}}

\newcommand{\msgs}[1][]{\Msgs[#1]}
\newcommand{\Msgs}[1][]{\mathit{Msgs}^{#1}}
\newcommand{\msgsphi}[3]{{\Msgs}_{#1}^{#2 \rightarrow #3}}

\newcommand{\glob}[2]{{\mathop{\mathit{global}}\left(\ifstrempty{#1}{\dots}{#1},\ifstrempty{#2}{\dots}{#2}\right)}}
\newcommand{\local}[1]{\mathop{\mathit{local}}\ifstrempty{#1}{}{\left(#1\right)}}
\newcommand{\external}[2]{{\mathop{\mathit{ext}}\left(\ifstrempty{#1}{\dots}{#1},\ifstrempty{#2}{\dots}{#2}\right)}}
\newcommand{\gsend}[4]{\mathop{\mathit{gsend}}(%
    \ifstrempty{#1}{\dots}{#1},%
    \ifstrempty{#2}{\dots}{#2},%
    \ifstrempty{#3}{\dots}{#3},%
    \ifstrempty{#4}{\dots}{#4}%
)}
\newcommand{\send}[2]{\mathop{\mathit{send}}(%
    \ifstrempty{#1}{\dots}{#1},%
    \ifstrempty{#2}{\dots}{#2}%
)}

\newcommand{\grecv}[4]{\mathop{\mathit{grecv}}(%
    \ifstrempty{#1}{\dots}{#1},%
    \ifstrempty{#2}{\dots}{#2},%
    \ifstrempty{#3}{\dots}{#3},%
    \ifstrempty{#4}{\dots}{#4}%
)}
\newcommand{\recv}[2]{\mathop{\mathit{recv}}(%
    \ifstrempty{#1}{\dots}{#1},%
    \ifstrempty{#2}{\dots}{#2}%
)}

\newcommand{\fakeof}[2]{{\mathop{\mathit{fake}}\left(#1,#2\right)}}
\newcommand{\localstates}[2][]{%
    {\mathscr{L}^{#1}_{#2}}%
}
\newcommand{\globalstates}{\mathscr{G}}
\newcommand{\globalinitialstates}[1][]{\mathscr{G}_{#1}(0)}
\newcommand{\localinitialstates}[1][]{\Omega_{#1}}

\newcommand{\failof}[1]{{\mathop{\mathit{fail}}\left(#1\right)}}
\newcommand{\sleep}[1]{{\mathop{\mathit{sleep}}\left(#1\right)}}
\newcommand{\hib}[1]{{\mathop{\mathit{hibernate}}\left(#1\right)}}

\newcommand{\tick}{\textbf{noop}}

\newcommand{\mistakefor}[2]{#1\mapsto#2}

\newcommand{\correct}[1]{correct_{#1}}
\newcommand{\faulty}[1]{faulty_{#1}}
\newcommand{\fake}[2][]{\mathit{fake}_{#1}\left(#2\right)}

\newcommand{\occurred}[2][]{\mathit{occurred}_{#1}(#2)}
\newcommand{\trueoccurred}[2][]{\overline{\mathit{occurred}}_{#1}(#2)}

\newcommand{\init}[2]{\overline{\mathit{init}}_{#1}(#2)}

\newcommand{\happened}[2][]{\mathit{happened}_{#1}(#2)}
\newcommand{\fhappened}[2][]{\mathit{fhappened}_{#1}(#2)}

\newcommand{\always}[1]{\square{#1}}

\newcommand{\START}{\operatorname{START}}
\newcommand{\FIRE}{\operatorname{FIRE}}

\newcommand{\envprotocol}[1]{P_{\epsilon}\ifstrempty{#1}{}{\left(#1\right)}}
\newcommand{\agprotocol}[2]{{P_{#1}\ifstrempty{#2}{}{\left(#2\right)}}}
\newcommand{\joinprotocol}[1]{P\ifstrempty{#1}{}{\left(#1\right)}}

\newcommand{\adversary}{adversary}

\newcommand{\alphae}[3][]{{%
    \alpha_{\epsilon_{#1}}^{#3}\ifstrempty{#2}{}{\left({#2}\right)}%
}}

\newcommand{\alphaag}[3]{{%
    \alpha_{#1}^{#3}\ifstrempty{#2}{}{\left(#2\right)}%
}}

\newcommand{\betae}[3][]{{%
    \beta_{\epsilon_{#1}}^{#3}\ifstrempty{#2}{}{\left({#2}\right)}%
}}

\newcommand{\betaag}[3]{{%
    \beta_{#1}^{#3}\ifstrempty{#2}{}{\left(#2\right)}%
}}

\newcommand{\filtere}[3][]{filter^{#1}_{\epsilon}\ifstrempty{#2}{}{\left(#2,#3\right)}}
\newcommand{\filterag}[4][]{filter^{#1}_{#2}\ifstrempty{#3}{}{\left(#3,#4\right)}}
\newcommand{\update}[2]{update\ifstrempty{#1}{}{\left(#1,#2\right)}}
\newcommand{\updatee}[2]{update_{\epsilon}\ifstrempty{#1}{}{\left(#1,#2\right)}}
\newcommand{\updateag}[4]{update_{#1}\ifstrempty{#2}{}{\left(#2,#3,#4\right)}}

\newcommand{\sigmaof}[1]{\sigma\ifstrempty{#1}{}{\bigl(#1\bigr)}}
\newcommand{\transition}[2][{\envprotocol{},\joinprotocol{}}]{\tau_{#1}\ifstrempty{#2}{}{\left({#2}\right)}}

\newcommand{\run}[3][]{r#1_{#2}\left(#3\right)}

\newcommand{\transitionExt}[3][{\envprotocol{},\joinprotocol{}}]{\tau^{#2}_{#1}\ifstrempty{#3}{}{\left({#3}\right)}}

\newcommand{\tauprotocol}[3]{\tau^{#1}_{{#2},{#3}}}

\newcommand{\Admissibility}[1][]{\Psi^{#1}}

\newcommand{\system}[1]{{R^{#1}}}

\newcommand{\pwrelation}[2][]{\sim^{#1}_{#2}}

\newcommand{\planguage}{\mathfrak{L}_{\mathfrak{g}}}
\newcommand{\intsys}{\I}

\newcommand{\I}{\mathcal{I}}
\newcommand{\prop}{\mathit{Prop}}

\newcommand{\dirobfag}{\mathsf{DirObBelFaultyAg}}
\newcommand{\notifbfag}{\mathscr{B}}

\newcommand{\dirnotifbfag}{\mathsf{DirNotifBelFaultyAg}}
\newcommand{\dirobmkf}{\mathsf{DirObMeKnowFaulty}}

\newcommand{\true}{\mathbf{true}}
\newcommand{\false}{\mathbf{false}}

\newcommand{\recvphiall}[3]{\widehat{Recv}_{#1}^{#3}(#2)}
\newcommand{\recvphi}[3]{Recv_{#1}^{#3}(#2)}
\newcommand{\agseq}{AgSeq}
\newcommand{\disjss}[1]{DisjSS^{#1}}

\newcommand{\ownsubsubsection}[1]{\medskip\noindent\textbf{#1}}

\newcommand{\proofarg}[1]{{\scriptsize\color{ForestGreen}(#1)}}

\usepackage{subcaption}
\usetikzlibrary{arrows,shapes,snakes,automata,backgrounds,petri}
\tikzstyle{m}=[circle, thin, draw,
minimum size=14mm,inner sep=3pt]
\tikzstyle{nod}= [circle, draw,inner sep=0pt, minimum size=0.5cm ]
\tikzset{
reflexive left/.style={->,loop,looseness=10,in=160,out=220},
reflexive right/.style={->,loop,looseness=10,in=20,out=320}
}

\title{A Sufficient Condition for Gaining Belief in Byzantine Fault-Tolerant Distributed Systems\thanks{Supported by Digital Modeling of Asynchronous Integrated Circuits (P32431-N30).}}
\author{Thomas Schl\"ogl
\institute{TU Wien, Vienna, Austria}
\email{tschloegl@ecs.tuwien.ac.at}
\and 
Ulrich Schmid
\institute{TU Wien, Vienna, Austria}
\email{s@ecs.tuwien.ac.at}
}
\def\authorrunning{T. Schl\"ogl and U. Schmid}
\def\titlerunning{A Sufficient Condition for Gaining Belief in Byzantine Fault-Tolerant Distributed Systems}

\begin{document}
\maketitle              

\setlength{\belowdisplayskip}{0pt} \setlength{\belowdisplayshortskip}{0pt}
\setlength{\abovedisplayskip}{0pt} \setlength{\abovedisplayshortskip}{0pt}

\begin{abstract}
Existing protocols for byzantine fault tolerant distributed systems usually rely on the correct agents' ability
to detect faulty agents and/or to detect the occurrence of some event or action on some 
correct agent. In this paper, we provide sufficient conditions that allow an agent to infer the
appropriate beliefs from its history, and a procedure that allows these conditions to be checked in finite time.
Our results thus provide essential stepping stones for developing efficient protocols and proving them correct.
\end{abstract}



\section{Introduction} \label{sec:intro}

At least since the ground-breaking work by Halpern and Moses \cite{HM90},
epistemic logic and interpreted runs and systems \cite{bookof4} are known 
as powerful tools for analyzing distributed systems. \emph{Distributed systems}
are multi-agent systems, where a set of $n\geq 2$ agents, each executing some protocol,
exchange messages in order to achieve some common goal.
In the \emph{interpreted runs and systems} framework,
the set of all possible runs $R$ (executions) 
of the agents in a system determines a set of Kripke models, formed by the evolution 
of the global state $r(t)$ in all runs $r\in I$ over time $t\in\mathbb{N}$.
Epistemic reasoning has been extended to \emph{fault-tolerant} distributed 
systems right from the beginning, albeit restricted to benign faulty agents,
i.e., agents that may only crash and/or drop messages 
\cite{moses1986programming,MT88,dwork1990knowledge,HM90}.

Actions performed by the agents when executing their protocol 
take place when they have accumulated specific epistemic knowledge. 
According to the pivotal \emph{Knowledge of Preconditions Principle}~\cite{Mos15TARK}, 
it is universally true that if $\varphi$ is a necessary condition for
an agent to take a certain action, then
$i$ may act only if $K_i \varphi$ is true. For example, in order
for agent~$i$ to decide on 0 in a binary fault-tolerant consensus 
algorithm \cite{LSP82} (where correct agents must
reach a common decision value based on local initial values),
it must know that some process has started with initial value 0, 
i.e., $K_i \varphi \equiv K_i(\exists j: x_j=0)$ holds true. Showing that agents act
without having attained $K_i \varphi$ for some necessary knowledge $\varphi$
is hence a very effective way for proving impossibilities. Conversely,
optimal distributed algorithms can be provided by letting agents act
as soon as $K_i \varphi$ for all the necessary knowledge has been 
established. One example are the crash-resilient \emph{unbeatable} consensus protocols 
introduced in \cite{CGM22:DC}, which are not just worst-case optimal,
but not even strictly dominated w.r.t.\ termination time by any other 
protocol in \emph{any} execution.

Epistemic reasoning has recently been extended to the analysis of \emph{byzantine} distributed systems 
\cite{KPSF19:FroCos,KPSF19:TARK,SSK20:PRIMA,FKS21:TARK} as well, where 
agents may not just crash or lose messages, but where they may also misbehave arbitrarily 
\cite{LSP82}. Solving a distributed computing problem in such systems
is much more difficult, and tighter constraints (e.g.\ on the maximum number $f$ of
faulty agents) are usually needed. For example, byzantine consensus can only be solved 
if $n \geq 3f+1$ \cite{LSP82}, whereas $n>f$ is sufficient for agents that may crash only 
\cite{CGM22:DC}.

When inspecting existing byzantine fault-tolerant protocols, one identifies
two basic tasks that usually need to be solved by every correct agent,
in some way or other: (1) detecting faulty agents, and
(2) detecting whether some correct agents are/were in a certain state.
We mentioned already a ``static'' example for (2), namely, finding out
whether some correct
agent started with initial value 0 in unbeatable consensus protocols
\cite{CGM22:DC}, which is also needed in byzantine-resilient protocols
like \cite{ST87:abc}. 
For a more dynamic example, we note that several existing byzantine fault-tolerant 
protocols employ the fundamental \emph{consistent broadcasting} (CB) primitive,
introduced in \cite{ST87}.
In particular, CB is used in fault-tolerant clock synchronization
\cite{ST87,WS09:DC,RS11:TCS}, in byzantine synchronous consensus
\cite{ST87:abc,DLS88}, and (in a slightly extended form) in the simulation
of crash-prone protocols in byzantine settings proposed in \cite{MTH14:STOC}. 
A variant of CB
has been studied epistemically in \cite{FKS21:TARK},
namely, \emph{firing rebels with relay} (FRR), which is the problem of
letting all correct agents execute an action $\FIRE$ in an all-or-nothing
fashion when sufficiently many agents know of an external $\START$ event.
It was shown that any correct protocol for implementing FRR (and, hence, CB) 
requires detecting whether $\START$ has occurred on some correct process.

Regarding an example for (1), we point out that it has been
shown by Kuznets et.~al.\ in 
\cite{KPSF19:FroCos} that it is impossible to
reliably detect whether some process is \emph{correct} in asynchronous byzantine
distributed systems, due to the possibility of a brain-in-a-vat scenario, whereas 
it is sometimes possible to detect that an agent is
\emph{faulty}. And indeed, the ability to (sometimes) 
reliably diagnose an actually byzantine 
faulty agent, using approaches like \cite{AR89}, has enabled the design of 
\emph{fault-detection, isolation and recovery} (FDIR) schemes \cite{PABB99} for 
high-reliability systems.

In this paper, we will provide sufficient conditions
that allow a correct agent $i$ to gain belief about a fact
$\varphi$ encoding (1) resp.\ (2) that is inherently local at some other correct
agent $j$. Using the \emph{belief modality} (also known as defeasible knowledge
\cite{MosSho93AI})
$B_i\varphi \equiv K_i(\correct{i} \to \varphi)$ that captures what is known
by agent~$i$ if it is correct, and the \emph{hope modality} $H_i\varphi \equiv 
\correct{i} \to B_i\varphi$ introduced in \cite{KPSF19:FroCos}, this can be succinctly
condensed into the following question: \emph{Under which conditions and by means
of which techniques can $B_iH_j\varphi$, which is the belief that $i$ obtains
by receiving a message from $j$ that claims (possibly wrongly) that $B_j\varphi$
holds, be lifted to $B_i\varphi$ in asynchronous byzantine systems?}
The crucial difference is that correct agent $i$
can infer something about $\varphi$ from the latter, but not from the former.
Note that it has been established in \cite[Thm.~15]{KPSF19:TARK} that
$B_i\varphi$ is indeed necessary for agent $i$ to achieve this.

\noindent
\emph{Detailed contributions:} For an asynchronous byzantine system with
weak communication assumptions, 
\begin{compactenum}
\item[(1)] we provide an algorithm that allows an agent $j$ to compute its belief
about the faultiness of the agents, based on both directly received obviously faulty
messages and on appropriate notifications from sufficiently many other agents recorded
in its local history,
\item[(2)] we provide a sufficient condition for agent $j$ to infer, also from
its local history, the belief that some event or action has occurred at a correct agent. 
\end{compactenum}
Our conditions are sufficient in the sense that if they hold, then the appropriate belief
can be obtained. Hence, the question about whether our conditions are also necessary 
might pop up. It is important to note, however, that the possibility of gaining belief
depends heavily on the actual properties of the system. For example, it will turn out
that the condition for (2) can be relaxed when a non-empty set of faulty agents 
is available, e.g.\ obtained via (1). The same is true if the actual system 
satisfies stronger communication assumptions. Consequently, we just focus on sufficient
conditions for communication
assumptions met by all asynchronous byzantine systems we are aware of.


\emph{Paper organization:} In \cref{sec:model}, we briefly introduce the
cornerstones of the byzantine modeling framework of \cite{KPSF19:FroCos,PKS19:TR}
needed for proving our results. \cref{sec:comm} provides our basic communication 
assumptions. \cref{sec:faults} and \cref{sec:events} contains our results
for detecting faulty agents (1) and the occurrence of events (2), respectively. 
Some conclusions and directions of future work are provided in \cref{sec:concl}.
Due to lack of space, additional technical details and all the proofs have been
relegated to an appendix.

\section{The Basic Model} \label{sec:model}

Since this paper uses the framework of~\cite{KPSF19:FroCos}, we restate
the core terms and aspects needed for our results.
 
There is a finite set~$\agents=\{1,\dots,n\}$ (for $n \ge 2$) of \textbf{agents}, who do not have access to a global clock and execute a possibly non-deterministic joint \textbf{protocol}.
In such a protocol, agents can perform \textbf{actions}, e.g.,~send \textbf{messages} $\mu\in\msgs$, and witness \textbf{events}, in particular,~message deliveries: the action of sending a copy (numbered $k$) of a message $\mu \in \msgs$ to an agent $j\in\agents$ in a protocol is denoted by $\send{j}{\mu_k}$, whereas a receipt of such a message from  $i\in\agents$ is recorded locally as $\recv{i}{\mu}$. 
The set of all \textbf{actions} (\textbf{events}) available to an agent $i\in \agents$ is denoted by $\actions[i]$ ($\events[i]$), subsumed as \textbf{haps} $\haps[i] \ce \actions[i]\sqcup\events[i]$, with $\actions \ce \bigcup_{i\in \agents}\actions[i]$, $\events \ce \bigcup_{i\in \agents}\events[i]$, and $\haps \ce \actions \sqcup \events$. 

The other main player in \cite{KPSF19:FroCos} is the \textbf{environment} $\epsilon$, which takes care of scheduling haps, failing agents, and resolving non-deterministic choices in the joint protocol.
Since the notation above only describes the local view of agents, there is also a \textbf{global} syntactic representation of each hap, which is only available to the environment and contains additional information (regarding the time of a hap, a distinction whether a hap occurred in a correct or byzantine way, etc.).
One distinguishes the 
sets of global events $\gevents[i] \ce \gtrueevents[i] \sqcup \bevents[i] \sqcup \sevents[i]$ of agent $i$, for a
correct agent (signified by the horizontal bar), a byzantine faulty agent, or a system event as explained below. Regarding
global actions, one distinguishes correct actions $\gtrueactions[i]$ and faulty actions 
$\fakeof{i}{\mistakefor{A}{A'}}$, where the agent actually performs $A$ but claims to have performed $A'$. Finally,
$\gevents\ce \bigsqcup_{i\in\agents} \gevents[i]$, $\ghaps \ce \gevents \sqcup \gactions$.
Generally, horizontal bars signify phenomena that are correct, as contrasted by those that may be correct or byzantine.

The model is based on discrete time, of arbitrarily fine resolution, with time domain $t\in \bbbt\ce\bbbn=\{0,1,\dots\}$.
All haps taking place after a \textbf{timestamp} $t\in\bbbt$ and no later than $t+1$ are grouped into a \textbf{round} denoted $t$\textonehalf{} and treated as happening simultaneously.
In order to prevent agents from inferring the global time by counting rounds, agents are generally unaware of a round, unless they perceive an event or are prompted to act by the environment.
The latter is accomplished by special system events $go(i)$, which are complemented by two more system events for faulty agents: $\sleep{i}$ and $\hib{i}$ signify a failure to activate the agent's protocol and differ in that the latter does not even wake up the agent. 
None of the \textbf{system events} $\sevents[i] \ce \{go(i), \sleep{i},\hib{i}\}$ is directly observable by agents.

Events and actions that can occur in each round, if enabled by $go(i)$, are determined by the protocols for agents and the environment, with non-deter\-min\-istic choices resolved by the \textbf{adversary} that is considered part of the environment.
A \textbf{run} $r$ is a function mapping a point in time $t$ to an $n+1$ tuple, consisting of the environment's history and local histories $r(t) = (r_{\epsilon}(t),r_1(t),\dots,r_n(t))$ representing the state of the whole system (\textbf{global state}) at that time $t$. The set of all global states is denoted by~$\globalstates$.
The \textbf{environment's history}~$r_\epsilon(t) \in \localstates{\epsilon}$ is a sequence of all haps that happened, in contrast to the local histories faithfully recorded in the global format.
Accordingly, $r_\epsilon(t+1) = X \circ r_\epsilon(t)$ for the set~$X \subseteq \ghaps$ of all haps from round~$t$\textonehalf{}, where $\circ$~stands for concatenation.
Agent $i$'s local view of the system after round $t$\textonehalf{}, i.e., its share of the global state $h=r(t) \in \globalstates$,
is recorded in $i$'s \textbf{local state}~$r_i(t+1)\in \localstates{i}$, also called $i$'s \textbf{local history}, sometimes denoted $h_i$.
$r_i(0)\in\localinitialstates[i]$ are the \textbf{initial local states}, with $\globalinitialstates \ce \prod_{i\in \agents}\localinitialstates[i]$. 
If a round contains neither  $go(i)$ nor any event to be recorded in  $i$'s local history, then the history $r_i(t+1)=r_i(t)$ remains unchanged, denying the agent knowledge that the round just passed. 
Otherwise, the agent performs actions from its protocol $\agprotocol{i}{\run{i}{t}}\subseteq 2^{\actions[i]}$ and updates its history
$r_i(t+1) = X \circ r_i(t)$, for the set $X \subseteq \haps[i]$ of all actions and events perceived by~$i$ in round~$t$\textonehalf{}.  
The sets $\betae[i]{r}{t}$, $\betaag{i}{r}{t}$ denote the sets of events and the set of actions respectively happening in round $t$\textonehalf{} in global format.
For some hap $o$ we write $o \in r_i(t)$ if there exists a round $t' \le t$, where $o$ was appended to $i$'s local history.
Consequently, the local history~$r_i(t)=h_i=(h_i(|h_i|),h_i(|h_i|-1),\dots,h_i(0))$ is the sequence of all haps $h_i(k)$ perceived by $i$ in the $k$-th round
it was \textbf{active} in.
\looseness=-1

The exact updating procedure is the result of a complex state transition consisting of several phases, 
described in detail in \cref{sec:protocols}, which are grouped into a \textbf{transition template} $\tau$ that yields a transition relation $\transitionExt{}{}$ for any joint and environment protocol $P$ and $P_\epsilon$. The set $\system{}$ of all \textbf{transitional runs} are all runs that can be generated from some set of initial states $\globalinitialstates$ via some transition template $\tau$.
\looseness=-1

Proving the correctness of a protocol for solving a certain distributed computing problem boils down to studying the set of runs that can be generated.
As \textbf{liveness properties} cannot be ensured on a round-by-round basis, they are enforced by restricting the allowable set of runs via \textbf{admissibility conditions} $\Psi$, which are subsets of the set $\system{}$ of all transitional runs. 
A \textbf{context} $\gamma=(\envprotocol{},\globalinitialstates,\tau,\Psi)$ consists of an environment's protocol $\envprotocol{}$, a set of global initial states $\globalinitialstates$, a transition template $\tau$, and an admissibility condition $\Psi$.
For a joint protocol $\joinprotocol{}$, we call $\chi=(\gamma,\joinprotocol{})$ an \textbf{agent context}.
The set of all $\chi$-consistent runs is denoted by $\system{\chi}$ that is the set of all transitional runs starting with initial states from $\globalinitialstates$ and transitioning via $\transitionExt{}{}$ (both from $\chi$).
The set of all agent contexts we denote by $\extension$, where $\extension[B] \subseteq \extension$ consists of all byzantine asynchronous agent contexts, with transition template $\tauprotocol{B}{\envprotocol{}}{\joinprotocol{}}$, and $\extension[B_f] \subseteq \extension[B]$ consists of all byzantine asynchronous agent contexts, where at most $f$ agents can become faulty, with transition template $\tauprotocol{B_f}{\envprotocol{}}{\joinprotocol{}}$.

\begin{toappendix}
\subsection{Global haps and faults.}\label{sec:global} As already mentioned, there is a global version of every $\haps$ that provides
additional information that is only accessible to the environment.
Among it is the timestamp $t$
For correct action $a\in\actions[i]$, as initiated by agent $i$ in the local format, the one-to-one function $\glob{i,t}{a}$ gives the global version.
Timestamps are especially crucial for proper message processing with
$
	\glob{i,t}{\send{j}{\mu_k}} \ce \gsend{i}{j}{\mu}{id(i,j,\mu,k,t)}
$ 
for some one-to-one function $id \colon \agents \times \agents \times \msgs \times \bbbn \times \bbbt \to \bbbn$ that assigns each sent message a unique \textbf{global message identifier} (GMI). 
These GMIs enable the direct linking of send actions to their corresponding delivery events, most importantly used to ensure that only sent messages can be delivered (causality).

Unlike correct actions, correct events witnessed by agent $i$ are generated by the environment $\epsilon$, hence are already produced in the global format $\gtrueevents[i]$. 
For each correct event $E \in \gtrueevents[i]$, we use a faulty counterpart $\fakeof{i}{E}$ and will make sure that agent $i$ cannot distinguish between the two. 
An important type of correct global events is delivery $\grecv{j}{i}{\mu}{id}\in \gtrueevents[i]$ of message $\mu$ with GMI $id \in \bbbn$ sent from agent $i$ to agent $j$. 
The GMI must be a part of the global format (especially for ensuring causality) but cannot be part of the local format because it contains information about the time of sending, which should not be accessible to agents.
The stripping of this information before updating local histories is achieved by the function 
$
	\mathit{local} \colon  \gtruethings \longrightarrow  \haps
$
converting \textbf{correct} haps from the global into the local formats for the respective agents in such a way that $\mathit{local}$ reverses $\mathit{global}$, i.e., $\mathit{local}\bigl(\glob{i,t}{a}\bigr) \ce a$, in particular, $\mathit{local}{\bigl(\grecv{i}{j}{\mu}{id}\bigr)} \ce \recv{j}{\mu}$. 

Faulty actions are modeled as byzantine events of the form $\fakeof{i}{\mistakefor{A}{A'}}$ where $A, A' \in \gtrueactions[i] \sqcup\{\tick\}$ for a special \textbf{non-action} $\tick$ in global format. 
These byzantine events are controlled by the environment and correspond to an agent violating its protocol by performing the action $A$, while recording in its local history that it either performs $a' = \mathit{local}(A')\in \actions[i]$ if $A' \in \gactions[i]$ or does nothing if $A' = \tick$. 


\subsection{Protocols, state transitions and runs.}\label{sec:protocols} The events and actions that occur in each round are determined by protocols (for agents and the environment) and non-determinism (adversary).
Agent $i$'s \textbf{protocol} 
$\agprotocol{i}{} \colon \localstates{i} \to 2^{2^{\actions[i]}}\setminus\{\varnothing\}$ 
provides a range $\agprotocol{i}{r_i(t)}$ of sets of actions based on $i$'s current local state~$r_i(t) \in \localstates{i}$ at time $t$ in run $r$, from which the adversary non-deterministically picks one.
Similarly the environment provides a range of (correct, byzantine, and system) events via its protocol 
$\envprotocol{} \colon \bbbt \to 2^{2^{\gevents}}\setminus \{\varnothing\}$, 
which depends on a timestamp~$t\in\bbbt$ but \textbf{not} on the current state, in order to maintain its impartiality.
%
It is required that all events of round~$t$\textonehalf{} be mutually compatible at time~$t$, called $t$-coherent according
to \cref{def:incomp}. The set of all global states is denoted by~$\globalstates$.\looseness=-1

\begin{definition}[Coherent events]
\label{def:incomp}
Let $t \in \mathbb{N}$ be a timestamp.
A set $S \subset \gevents$ of events is called \textbf{$t$-coherent} if it satisfies the following conditions:
\begin{compactenum}
\item for any $\fakeof{i}{\mistakefor{\gsend{i}{j}{\mu}{id}}{A}}\in S$, the GMI $id = id(i,j,\mu,k,t)$ for some $k\in \mathbb{N}$;
\item for any $i \in \agents$ at most one of $go(i)$, $\sleep{i}$, and $\hib{i}$ is present in $S$;
\item for any $i \in \agents$ and any $e \in Ext_i$   at most one of $\external{i}{e}$ and  $\fakeof{i}{\external{i}{e}}$ is present in $S$;
\item for any $\grecv{i}{j}{\mu}{id_1} \in S$, no event of the form $\fakeof{i}{\grecv{i}{j}{\mu}{id_2}}$ belongs to $S$ for any $id_2 \in \mathbb{N}$;
\item for any $\fakeof{i}{\grecv{i}{j}{\mu}{id_1}} \in S$, no event of the form $\grecv{i}{j}{\mu}{id_2}$ belongs to $S$ for any $id_2 \in \mathbb{N}$.
\end{compactenum}
%
\end{definition}


Given the \textbf{joint protocol}~$P\ce(P_1,\dots,P_n)$ and the environment's protocol~$P_\epsilon$, we focus on \textbf{$\transitionExt{}{}$-transitional runs}~$r$ that result from following these protocols and are built according to a \textbf{transition relation}~$\transitionExt{}{}\subseteq \globalstates \times \globalstates$.
Each such transitional run begins in some initial global state $r(0)\in\globalinitialstates$ and progresses, satisfying $(\run{}{t},\run{}{t+1}) \in \transitionExt{}{}$ for each timestamp $t\in\bbbt$. 

\begin{figure}[t]
    \begin{center}
        \scalebox{0.6}{
            \input{transition_function_WoLLIC.tex}
        }
        \caption{The evolution of states in round $t.5$ (from timestamp $t\in\mathbb{N}$ to $t+1$) inside a run $r$ constructed according to the transition function $\transition{}$. Different communication models require changes to the filtering functions $\filtere{}{}{}$ and $\filterag{i}{}{}$.}
        \label{fig:trans_rel}
    \end{center}
\end{figure}
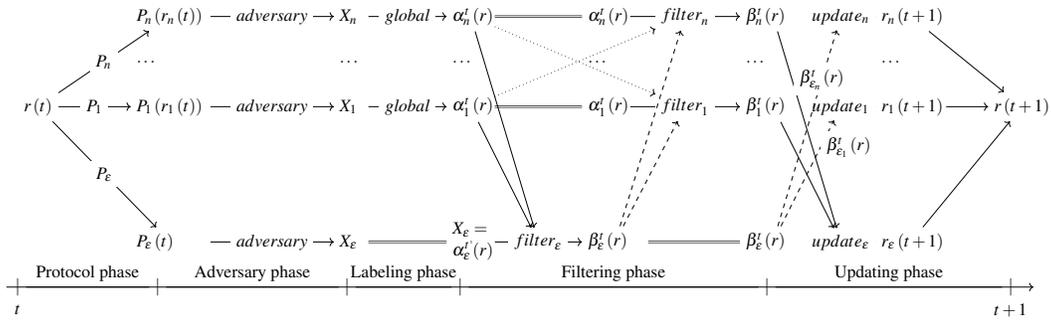

The transition relation $\transitionExt{}{}$ consisting of  five consecutive  phases is illustrated in \cref{fig:trans_rel}
and works as follows:

In the \textbf{protocol phase}, a range 
$\envprotocol{t} \subset 2^{\gevents}$ 
of $t$-coherent sets of events is determined by the environment's protocol $P_\epsilon$.
Similarly for each $i\in\agents$, a range 
$\agprotocol{i}{\run{i}{t}}\subseteq 2^{\actions[i]}$ 
of sets of $i$'s actions is determined by the joint protocol~$P$.\looseness=-1

In the \textbf{adversary phase}, the adversary non-deterministically chooses a set $X_\epsilon \in \envprotocol{t}$ and one set $X_i \in \agprotocol{i}{\run{i}{t}}$ for each $i\in\agents$.

In the \textbf{labeling phase}, actions in the sets $X_i$ are translated into the global format: $\alphaag{i}{r}{t}\ce$ \\
$\{\glob{i,t}{a} \mid a \in X_i\}\subseteq \gtrueactions[i]$.

In the \textbf{filtering phase}, filter functions remove all unwanted or impossible attempted events from $\alphae{r}{t}\ce X_\epsilon$ and actions from $\alphaag{i}{r}{t}$. 
This is done in two stages:

\noindent
First, $\filtere{}{}$ filters out ``illegal'' events.
This filter will vary depending on the concrete system assumptions (in the byzantine asynchronous case, ``illegal'' constitutes receive events that violate causality).
The resulting set of events to actually occur in round $t$\textonehalf{} is $\betae{r}{t} \ce \filtere{\run{}{t}}{\alphae{r}{t}, \alphaag{1}{r}{t}, \dots, \alphaag{n}{r}{t}}$.
The byzantine asynchronous filter (ensuring causality) is denoted by
$\filtere[B]{h}{X_\epsilon, X_1, \dots, X_n}$
and the byzantine asynchronous at-most-f-faulty-agents filter, which both ensures causality and removes all byzantine events if as a result the total number of faulty agents were to exceed $f$ is denoted by \\
$\filtere[B_f]{h}{\,X_\epsilon,\, X_1,\, \dots,\, X_n}$.

\begin{definition} \label{def:std_ac_filter}
	The \textbf{standard action filter} $\filterag[B]{i}{X_1, \dots, X_n}{X_\epsilon}$ for $i \in \agents$ either removes all actions from $X_i$ when $go(i) \notin X_\epsilon$ or else leaves $X_i$ unchanged.\looseness=-1
\end{definition}
\noindent
Second, $\filterag[B]{i}{}{}$ for each $i$ returns the sets of actions to be actually performed by agents in round $t$\textonehalf{}, i.e., $\betaag{i}{r}{t} \ce \filterag[B]{i}{\alphaag{1}{r}{t}, \dots, \alphaag{n}{r}{t} }{\betae{r}{t}}$.
Note that $\betaag{i}{r}{t} \subseteq \alphaag{i}{r}{t} \subseteq \gtrueactions[i]$ and $\betae{r}{t} \subseteq \alphae{r}{t} \subset \gevents$.

In the \textbf{updating phase}, the events $\betae{r}{t}$ and actions $\betaag{i}{r}{t}$ are appended to the global history $r(t)$.
For each $i \in \agents$, all non-system events from $\betae[i]{r}{t} \ce \betae{r}{t}\cap\gevents[i]$ and all actions $\betaag{i}{r}{t}$ as \textbf{perceived} by the agent are appended in the local form to the local history $r_i(t)$.
Note the local history may remain unchanged if no events trigger an update.\looseness=-1
\begin{definition}[State update functions]\label{def:state-update}
        \index{${\updatee{}{},\updateag{i}{}{}{}}$}
           Given a global history  $h = (h_\epsilon, h_1, \dots, h_n) \in \globalstates$, a tuple of performed actions/events $X=(X_{\epsilon},X_1,\dots,X_n) \in 2^{\gevents} \times 2^{\gtrueactions[1]}\times \ldots \times 2^{\gtrueactions[n]}$,
we use the following abbreviation $X_{\epsilon_i} = X_{\epsilon} \cap \gevents[i]$ for 
each 
$i \in \agents$. 
               Agents $i$'s update function $\updateag{i}{}{}{} \colon \localstates{i} \times 2^{\gtrueactions[i]}\times2^{\gevents} \to \localstates{i}$
               outputs a new local history from $\localstates{i}$ based on $i$'s actions~$X_i$ and 
			   events~$X_{\epsilon}$ as follows:
               \begin{equation}
               \label{eq:update_agent}
                   \updateag{i}{h_i}{X_i}{X_\epsilon} \ce 
                   \begin{cases}
                   h_i & 
                       \text{if $\sigma(X_{\epsilon_i})=\varnothing$ and }
                      go(i) \not\in X_\epsilon \\
                       \Bigl[\sigmaof{ X_{\epsilon_i}\sqcup X_i} \Bigr] \circ h_i & \text{ otherwise }
                   \end{cases}
               \end{equation}
			   where $\sigma(X)$ removes all system events $\sevents$ from $X$ and afterwards invokes $\local{}$ on the resulting set.
        Similarly, the environment's state update function $\updatee{}{} \colon \localstates{\epsilon} \times \left(2^{\gevents} \times 2^{\gtrueactions[1]}\times \ldots \right.$\\
		$\left.\times 2^{\gtrueactions[n]}\right) \to \localstates{\epsilon}$ outputs a new state of the environment based on $X_{\epsilon}$:      
        \begin{equation}
        \label{eq:update_env}
        \updatee{h_\epsilon}{X} \ce (X_{\epsilon} \sqcup X_1 \sqcup \ldots \sqcup X_n) \colon h_\epsilon 
        \end{equation} 
        Thus, the global state is modified as follows:
\begin{align} \label{eq:update_global}
	\update{h}{X} &\ce \Bigl( \updatee{h_\epsilon}{X}, \updateag{1}{h_1}{X_1}{X_\epsilon}, \dots, \updateag{n}{h_n}{X_n}{X_\epsilon} \Bigr) \text{ and}\\
	\run{\epsilon}{t+1} &\ce \updatee{\run{\epsilon}{t}}{\quad\betae{r}{t},\quad \betaag{1}{r}{t},\quad \dots,\quad \betaag{n}{r}{t}} \\
	\run{i}{t+1} &\ce \updateag{i}{\run{i}{t}}{\quad\betaag{i}{r}{t}}{\quad\betae{r}{t}}. \label{eq:run_trans_ag}
\end{align}
\end{definition}
The operations in the phases 2--5 (adversary, labeling, filtering and updating phase) are grouped into a \textbf{transition template} $\tau$ that yields a transition relation $\transitionExt{}{}$ for any joint and environment protocol $P$ and $P_\epsilon$.
Particularly, we denote as~$\tau^B$ the transition template utilizing $\filtere[B]{}{}$ and $\filterag[B]{i}{}{}$ (for all $i \in \agents$).\looseness=-1

\end{toappendix}

\ownsubsubsection{Epistemics.}\label{page:epistemics} \cite{KPSF19:FroCos} defines interpreted systems in this framework as Kripke models for multi-agent distributed environments \cite{bookof4}. 
The states in such a Kripke model are given by global histories $r(t')\in \globalstates$ for runs $r \in \system{\chi}$ given some agent context $\chi$ and timestamps $t' \in \bbbt$.
A \textbf{valuation function} 
$\pi \colon \prop \to 2^{\globalstates}$ 
determines states where an atomic proposition from $\prop$ is true.
This determination is arbitrary except for a small set of \textbf{designated atomic propositions}:
For $\fevents[i]\ce \bevents[i]\sqcup\{\sleep{i},\hib{i}\}$, $i\in\agents$, $o \in \haps[i]$, and $t\in \bbbt$ such that $t \leq t'$,
\begin{compactitem} 
\label{page:occurred}
	\item $\correct{(i,t)}$ is true at $r(t')$ if{f} no faulty event happened to $i$ by timestamp $t$, i.e., no event from $\fevents[i]$ appears in $r_\epsilon(t)$,
	\item $\correct{i}$ is true at $r(t')$ if{f} no faulty event happened to $i$ yet, i.e., no event from $\fevents[i]$ appears in $r_\epsilon(t')$,
	\item $\faulty{i}$ is true if{f} $\neg\correct{i}$ is and $\faulty{(i,t)}$ is true if{f} $\neg\correct{(i,t)}$ is,
	\item $\fake[(i,t)]{o}$ is true at $r(t')$ if{f} $i$ has a \textbf{faulty} reason to believe  that $o\in\haps[i]$ occurred in round  $(t-1)$\textonehalf{}, i.e., $o \in r_i(t)$ because (at least in part) of some $O \in \bevents[i] \cap \betae[i]{r}{t-1}$ (see \cref{sec:protocols}),
	\item $\trueoccurred[(i,t)]{o}$ is true at $r(t')$ if{f} $i$ has a \textbf{correct} reason to believe $o\in\haps[i]$ occurred in round  $(t-1)$\textonehalf{}, i.e., $o \in r_i(t)$ because (at least in part) of $O \in (\gtrueevents[i] \cap \betae[i]{r}{t-1}) \sqcup  \betaag{i}{r}{t-1}$ (see \cref{sec:protocols}),
	\item $\trueoccurred[i]{o}$ is true at $r(t')$ if{f} at least one of $\trueoccurred[(i,m)]{o}$ for $1 \leq m \leq t'$ is; also $\trueoccurred{o}\ce \bigvee_{i\in\agents} \trueoccurred[i]{o}$,
	\item $\occurred[i]{o}$ is true at $r(t')$ if{f} either $\trueoccurred[i]{o}$ is or at least one of $\fake[(i,m)]{o}$  for $1 \leq m \leq t'$ is,
	\item $\happened[i]{a}$ is true at $r(t')$ for action $a \in \actions[i]$ if{f} there exists a global action $A$, where $a \in \local{A}$ s.t. $A \in r_\epsilon(t'-1)$ or $\fakeof{i}{\mistakefor{A}{A'}} \in r_\epsilon(t'-1)$,
	\item $\fhappened[i]{a}$ is true at $r(t')$ for action $a \in \actions[i]$ if{f} there exists a global action $A$, where $a \in \local{A}$ s.t. $\fakeof{i}{\mistakefor{A}{A'}} \in r_\epsilon(t'-1)$,
	\item $\init{i}{\lambda_0}$ is true at $r(t')$ for initial state $\lambda_0 \in \localinitialstates[i]$ (see \cref{sec:protocols}) if{f} $r_i(0) = \lambda_0$. Note that all agents are still correct in any initial state.
\end{compactitem}

The following terms are used to categorize agent faults caused by the environment's 
protocol~$\envprotocol{}$: agent~$i \in \agents$ is
\emph{fallible} if for any $X \in \envprotocol{t}$, $X \cup  \{\failof{i}\}  \in \envprotocol{t}$;
\emph{correctable} if  $X \in \envprotocol{t}$ implies that $X \setminus \fevents[i] \in \envprotocol{t}$;
\emph{delayable} if     $X \in \envprotocol{t}$ implies $X \setminus \gevents[i]  \in \envprotocol{t}$;
\emph{gullible} if $X \in \envprotocol{t} $ implies that, for any $Y \subseteq  \fevents[i]$,  the set
   $Y \sqcup (X \setminus \gevents[i])  \in \envprotocol{t}$ whenever it is $t$-coherent (consists of mutually
compatible events only, see \cref{def:incomp}).
Informally, fallible agents can be branded byzantine at any time;
correctable agents can always be made correct for the round by removing all their byzantine events; 
delayable agents can always be forced to skip a round completely (which does not make them faulty);  
gullible agents can exhibit any faults in place of correct events. 
Common types of faults, e.g.,~crash or omission failures, can be obtained by restricting allowable sets~$Y$ in the definition of gullible agents.

An \textbf{interpreted system} is a pair $\I = (\system{\chi}, \pi)$.
The following BNF defines the \textbf{epistemic language} $\planguage$ considered throughout this paper, for $p \in \prop$ and $i\in\agents$: $\varphi \cce p \mid \lnot \varphi \mid (\varphi \land \varphi) \mid K_i \varphi \mid \always{\varphi}$ (other Boolean connectives are defined as usual). We also use belief $B_{i} \varphi \ce K_{i} (\correct{i} \rightarrow \varphi)$ and hope $H_{i} \varphi \ce \correct{i} \to B_i \varphi$ as introduced in \cite{KPSF19:FroCos} and axiomatized in \cite{Fru19:ESSLLI, DFK22:AIML}.
The interpreted systems semantics  is defined as usual with global states  $r(t)$ and $r'(t')$ indistinguishable for  $i$ if{f} $r_i(t)=r'_i(t')$.
Semantics for temporal operator $\always{}$ we define as $(\intsys, r, t) \models \always{\varphi} \quad \text{if{f}} \quad (\forall t' \ge t)\ (\intsys, r, t') \models \varphi$.




\section{Basics of Fault-Tolerant Communication}
\label{sec:comm}

In this section, we introduce some basic notation and assumptions needed
for fault-tolerant communication. Throughout
this paper, we consider non-excluding agent contexts where all agents are
fallible, gullible, correctable, and delayable \cite{KPSF19:TARK}, which models
asynchronous byzantine distributed systems.

\begin{definition} \label{def:phi_msgs}
	For the interpreted system $\intsys = (\system{\chi}, \pi)$, formula $\varphi$ from $\planguage$, and agents $i,j \in \agents$, we define a set of ``trustworthy'' messages $\msgsphi{\varphi}{i}{j} \subseteq \msgs$ that an agent $i$ sends to $j$ only if $i$ believes $\varphi$,
	\begin{equation}
			\mu \in \msgsphi{\varphi}{i}{j} \quad \Longleftrightarrow \quad \big((\forall r \in \system{\chi})(\forall t \in \mathbb{N})(\forall D \in P_i(r_i(t)))\ \send{j}{\mu} \in D \ \Rightarrow \ (\intsys, r, t) \models B_i \varphi\big).
	\end{equation}
\end{definition}

%

\begin{definition} \label{def:persistence}
	We call a formula $\varphi$ \emph{persistent} in the interpreted system $\intsys = (\system{\chi}, \pi)$
if, once true, $\varphi$ never becomes false again:
	\begin{equation} \label{eq:persistence}
		(\forall r \in \system{\chi})(\forall t \in \mathbb{N})(\forall t' > t)\ (\intsys, r, t) \models \varphi \ \Rightarrow \ (\intsys, r, t') \models \varphi.
	\end{equation}
\end{definition}


Persistent formulas have several useful properties, which are easy to prove:\footnote{Most of our proofs have been relegated
to the appendix.}

\begin{lemma} \label{lem:faulty_occ_pers}
	For any agent context $\chi \in \extension[B]$, agent $i \in \agents$, natural number $k \in \mathbb{N}$, action or event $o \in \haps$ and $\lambda_0 \in \localinitialstates[i]$, the formulas $\faulty{i}$, $\occurred[i]{o}$, $\trueoccurred[i]{o}$ and $\init{i}{\lambda_0}$ 
are persistent.
\end{lemma}

\begin{lemma} \label{lem:cor_phi_pers}
	If, for an agent context $\chi$, formula $\varphi$ is persistent, so is $\correct{i} \rightarrow \varphi$ for any agent $i \in \agents$.
\end{lemma}

\begin{lemma} \label{lem:pers}
	If, for an agent context $\chi$, a formula $\varphi$ is persistent, so is $K_i \varphi$, $B_i \varphi$, $H_i \varphi$ for any agent $i \in \agents$.
\end{lemma}

\begin{lemma} \label{lem:conj_disj_pers}
	If, for an agent context $\chi$, formulas $\varphi, \psi$ are persistent, so is $\varphi \wedge \psi$ and $\varphi \vee \psi$.
\end{lemma}

The following essential lemma states what an agent $i$ can infer epistemically from
receiving a trustworthy message referring to a persistent formula $\varphi$ from agent $j$.
As explained in \cite[Remark~11]{FKS21:TARK}, in the case of $j$ being faulty, agent $i$
need not share its reality with agent $j$.
Consequently, agent $i$ cannot infer $B_i\varphi$ just from receiving such a message.

\begin{lemmarep} \label{lem:recv_BH}
	For agent context $\chi \in \extension[B]$, interpreted system $\intsys = (\system{\chi}, \pi)$, run $r \in \system{\chi}$, $t \in \mathbb{N}$, agents $i,j \in \agents$, persistent formula $\varphi$, and trustworthy message $\mu \in \msgsphi{\varphi}{j}{i}$
	\begin{equation}
		\begin{aligned}
		\recv{j}{\mu} \in r_i(t) \quad \Rightarrow \quad (\intsys, r, t) \models B_iH_j \varphi.
		\end{aligned}
	\end{equation}
\end{lemmarep}
\begin{proof}
	\begin{compactenum}
		\item $\mu \in \msgsphi{\varphi}{j}{i}$ \proofarg{by assumption} \label{it:thm_recv_BH_1}
		\item $\recv{j}{\mu} \in r_i(t)$ \proofarg{by assumption} \label{it:thm_recv_BH_2}
		\item $(\intsys, r, t) \not\models B_i H_j \varphi$ \proofarg{by assumption} \label{it:thm_recv_BH_3}
		\item $\widehat{r} \in \system{\chi} \text{ and } \widehat{t} \in \mathbb{N} \text{ and } r(t) \pwrelation{i} \widehat{r}(\widehat{t}) \text{ and } (\intsys, \widehat{r}, \widehat{t}) \models \correct{i} \ \wedge \ \neg H_j \varphi$ \proofarg{from line \eqref{it:thm_recv_BH_3}, by sem. of $B_i$, exist. inst.} \label{it:thm_recv_BH_4}
		\item $(\intsys, \widehat{r}, \widehat{t}) \models \occurred[i]{\recv{j}{\mu}}$ \proofarg{from lines \eqref{it:thm_recv_BH_2}, \eqref{it:thm_recv_BH_4}, by def. of $\pwrelation{i}$, sem. of $\occurred{}$, ``and''} \label{it:thm_recv_BH_5}
		\item
		\begin{compactenum}
			\item $(\intsys, \widehat{r}, \widehat{t}) \models \fake[i]{\recv{j}{\mu}}$ \proofarg{from line \eqref{it:thm_recv_BH_5}, by sem. of $\occurred[i]{}$, $\fake[i]{}$} \label{it:thm_recv_BH_6.1}
			\item $(\intsys, \widehat{r}, \widehat{t}) \models \faulty{i}$ \proofarg{from line \eqref{it:thm_recv_BH_6.1}, by sem. of $\fake[i]{}$} \label{it:thm_recv_BH_6.2}
			\item contradiction! \proofarg{from lines \eqref{it:thm_recv_BH_4}, \eqref{it:thm_recv_BH_6.2}, by sem. of ``and''} \label{it:thm_recv_BH_6.3}
		\end{compactenum}
		\item
		\begin{compactenum}
			\item $(\intsys, \widehat{r}, \widehat{t}) \models \trueoccurred[i]{\recv{j}{\mu}}$ \proofarg{from line \eqref{it:thm_recv_BH_5}, by sem. of $\occurred[i]{}$, $\trueoccurred[i]{}$} \label{it:thm_recv_BH_7.1}
			\item $(\intsys, \widehat{r}, \widehat{t}) \models \happened[j]{\send{i}{\mu}}$ \proofarg{from line \eqref{it:thm_recv_BH_7.1}, by def. of $\filtere[B]{}{}$} \label{it:thm_recv_BH_7.2}
			\item $(\intsys, \widehat{r}, \widehat{t}) \models \correct{i} \wedge \correct{j} \wedge \neg B_j \varphi$ \proofarg{from line \eqref{it:thm_recv_BH_4}, by sem. of $\neg H_j$} \label{it:thm_recv_BH_7.3}
			\item 
			\begin{compactenum}
				\item $(\intsys, \widehat{r}, \widehat{t}) \models \fhappened[j]{\send{i}{\mu}}$ \proofarg{from line \eqref{it:thm_recv_BH_7.2}, by sem. of $\happened[i]{}$, $\fhappened[i]{}$} \label{it:thm_recv_BH_7.4.1}
				\item $(\intsys, \widehat{r}, \widehat{t}) \models \faulty{j}$ \proofarg{from line \eqref{it:thm_recv_BH_7.4.1}, by sem. of $\fhappened[j]{}$} \label{it:thm_recv_BH_7.4.2}
				\item contradiction! \proofarg{from lines \eqref{it:thm_recv_BH_7.3}, \eqref{it:thm_recv_BH_7.4.2}, by sem. of $\wedge$} \label{it:thm_recv_BH_7.4.3}
			\end{compactenum}
			\item
			\begin{compactenum}
				\item $(\intsys, \widehat{r}, \widehat{t}) \models \trueoccurred[j]{\send{i}{\mu}}$ \proofarg{from line \eqref{it:thm_recv_BH_7.2}, by sem. of $\happened[j]{}$} \label{it:thm_recv_BH_7.5.1}
				\item $(\forall r \in \system{\chi})(\forall t \in \mathbb{N})(\forall D \in P_j(r_j(t)))\ \send{i}{\mu} \in D \Rightarrow (\intsys, r, t) \models B_j \varphi$ \proofarg{from line \eqref{it:thm_recv_BH_1}, by Definition \ref{def:phi_msgs}} \label{it:thm_recv_BH_7.5.2}
				\item $(\intsys, \widehat{r}, \widehat{t}) \models B_j \varphi$ \proofarg{from lines \eqref{it:thm_recv_BH_7.5.1}, \eqref{it:thm_recv_BH_7.5.2}, by univ. inst. sem. of $\trueoccurred[j]{}$, $\Rightarrow$, persistence of $\varphi$ and Lemma \ref{lem:pers}} \label{it:thm_recv_BH_7.5.3}
				\item contradiction! \proofarg{from lines \eqref{it:thm_recv_BH_7.3}, \eqref{it:thm_recv_BH_7.5.3}, by sem. of $\wedge$} \label{it:thm_recv_BH_7.5.4}
			\end{compactenum}
		\end{compactenum}
	\end{compactenum}
\end{proof}

In systems where not all agents can or want to communicate directly with all other
agents, messages need to travel multiple hops before they reach a desired recipient.
We therefore need to generalize \cref{lem:recv_BH} accordingly.

\begin{definition} \label{def:agSeq}
	We define the set of all finite agent sequences (including the empty sequence $\epsilon$), with repetitions allowed, as
	\begin{equation} \label{eq:agSeq}
		\agseq \ \ce \ \{(i_1, i_2, \ldots, i_k) \mid i_1, i_2, \ldots, i_k \in \agents\}.
	\end{equation}
\end{definition}

\begin{definition} \label{def:hopeChain}
	For agent sequence $\sigma \in \agseq$ and formula $\varphi$, we define the \emph{nested hope} 
$\overline{H}_{\sigma} \varphi$ for $\varphi$ as
	\begin{equation} \label{eq:hopeChain}
		\overline{H}_{\sigma} \varphi \ \ce \ H_{\pi_1 \sigma}H_{\pi_2 \sigma} \ldots H_{\pi_{|\sigma|}\sigma} \varphi,
	\end{equation}
	where $\pi_k\sigma$ denotes the application of the $k$th projection function to $\sigma$, and $\overline{H}_{\epsilon} \varphi = \varphi$.
\end{definition}
Note that the nested hope caused by $\sigma=(i_1, i_2, \ldots, i_k)$ goes from the origin $i_k$ to $i_1$.

\begin{definition} \label{def:recv_phi_all}
	For a persistent formula $\varphi$ in agent context $\chi$, run $r \in \system{\chi}$, time $t \in \mathbb{N}$ and agent $i \in \agents$, we define the set of agent sequences that lead to nested hope regarding $\varphi$ 
at agent $i$ as
	\begin{equation} \label{eq:recv_phi_all}
		\begin{aligned}
			\recvphiall{\varphi}{r_i(t)}{i} \ \ce \ \{\sigma \in \agseq \mid\ \mu \in \msgsphi{\overline{H}_{\overline{\sigma}}\varphi}{j}{i} \quad \text{and} \quad \sigma = (j) \circ \overline{\sigma} \quad \text{and} \quad \recv{j}{\mu} \in r_i(t)\},
		\end{aligned}
	\end{equation}
	where $\circ$ denotes sequence concatenation, i.e., $(j) \circ (i_1, \ldots, i_k) = (j, i_1, \ldots, i_k)$.
The agent sequences $\sigma \in \recvphiall{\varphi}{r_i(t)}{i}$ will be called \emph{hope chains}.
\end{definition}

With these preparations, we can generalize \cref{lem:recv_BH} to 
arbitrary hope chains as follows:
\begin{corollary} \label{cor:recv_phi_all_BH}
	For agent context $\chi \in \extension[B]$, interpreted system $\intsys = (\system{\chi}, \pi)$, run $r \in \system{\chi}$, $t \in \mathbb{N}$, agents $i,j \in \agents$, and persistent formula $\varphi$,
	\begin{equation} \label{eq:cor_recv_phi_all_BH}
		(\forall \sigma \in \recvphiall{\varphi}{r_i(t)}{i})\ (\intsys, r, t) \models B_i\overline{H}_\sigma \varphi.
	\end{equation}
\end{corollary}

The following two lemmas show that, for any hope chain in the set $\recvphiall{\varphi}{r_i(t)}{i}$ that contains a loop starting and ending in some agent $j\neq i$ in the actual communication chain, i.e., the agent sequence, the corresponding hope chain where the loop is replaced by a single instance of $j$ exists in $\recvphiall{\varphi}{r_i(t)}{i}$.
\begin{lemmarep} \label{lem:sigmaHC}
	For persistent $\varphi$, $\chi \in \extension[B]$, $\intsys = (\system{\chi}, \pi)$, $r \in \system{\chi}$,  $t \in \mathbb{N}$, $\widehat{\sigma} \in \recvphiall{\varphi}{r_i(t)}{i}$, $\sigma = (i) \circ \widehat{\sigma}$,
	\begin{equation}
		(\forall k \in [1, |\sigma|-1])\ (\intsys, r, t) \models \bigwedge\limits_{k' \in [1, k]} \correct{\pi_{k'} \sigma} \quad \Rightarrow \quad (\intsys, r, t) \models \overline{H}_{\pi_{k+1}\sigma \circ \ldots \circ \pi_{|\sigma|}\sigma} \varphi
	\end{equation}
\end{lemmarep}
\begin{proof}
	by induction over $k \in [1, |\sigma| - 1]$.
	\\
	\underline{Ind. hyp:} 
		$(\forall k \in [1, |\sigma|-1])\ (\intsys, r, t) \models \bigwedge\limits_{k' \in [1, k]} \correct{\pi_{k'} \sigma} \quad \Rightarrow \quad (\intsys, r, t) \models \overline{H}_{\pi_{k+1}\sigma \circ \ldots \circ \pi_{|\sigma|}\sigma} \varphi$
	\\
	\underline{Base case for $k = 1$:} by contradiction.
	\\
		\begin{compactenum}
			\item $(\intsys, r, t) \not\models \overline{H}_{\pi_2\sigma \circ \ldots \circ \pi_{|\sigma|}\sigma} \varphi$ \proofarg{by contr. assumption} \label{it:lem_sigmaHC_base1a}
			\item $(\intsys, r, t) \models \trueoccurred[i]{\recv{\pi_1 \sigma}{\mu}} \text{ and } \mu \in \msgsphi{\overline{H}_{\pi_3 \sigma \circ \ldots \circ \pi_{|\sigma|}\sigma}\varphi}{\pi_2 \sigma}{i}$ \proofarg{since $\widehat{\sigma} \in \recvphiall{\varphi}{r(t)}{i}$, $(\intsys, r, t) \models \correct{i}$, by Definition \ref{def:recv_phi_all}} \label{it:lem_sigmaHC_base1}
			\item $(\intsys, r, t) \models B_i\overline{H}_{\pi_2\sigma \circ \ldots \circ \pi_{|\sigma|}\sigma} \varphi$ \proofarg{from line \eqref{it:lem_sigmaHC_base1}, by Corollary \ref{cor:recv_phi_all_BH}} \label{it:lem_sigmaHC_base2}
			\item $(\intsys, r, t) \models \overline{H}_{\pi_2\sigma \circ \ldots \circ \pi_{|\sigma|}\sigma} \varphi$ \proofarg{from line \eqref{it:lem_sigmaHC_base2}, since $(\intsys, r, t) \models \correct{i}$} \label{it:lem_sigmaHC_base3}
			\item contradiction! \proofarg{from lines \eqref{it:lem_sigmaHC_base1a}, \eqref{it:lem_sigmaHC_base3}}
		\end{compactenum}
		\underline{Ind. step $k-1 \rightarrow k$ (for $k \in [2, |\sigma| - 1]$):} by contradiction.
		\\
		\begin{compactenum}
			\item $\widehat{\sigma} \in \recvphiall{\varphi}{r(t)}{i} \text{ and } \sigma = (i) \circ \widehat{\sigma}$ \proofarg{by contr. assumption} \label{it:lem_sigmaHC_step1}
			\item $(\intsys, r, t) \models \bigwedge\limits_{k' \in [1, k]} \correct{\pi_{k'} \sigma}$ \proofarg{by contr. assumption, sem. of $\Rightarrow$} \label{it:lem_sigmaHC_step2}
			\item $(\intsys, r, t) \not\models \overline{H}_{\pi_{k+1} \sigma \circ \ldots \circ \pi_{|\sigma|}\sigma} \varphi$ \proofarg{by contr. assumption} \label{it:lem_sigmaHC_step3}
			\item $(\intsys, r, t) \models \overline{H}_{\pi_{k} \sigma \circ \ldots \circ \pi_{|\sigma|}\sigma} \varphi$ \proofarg{by assumption of the ind. hyp. for $k-1$} \label{it:lem_sigmaHC_step4}
			\item $(\intsys, r, t) \models B_{\pi_k\sigma}\overline{H}_{\pi_{k+1} \sigma \circ \ldots \circ \pi_{|\sigma|}\sigma} \varphi$ \proofarg{from lines \eqref{it:lem_sigmaHC_step2}, \eqref{it:lem_sigmaHC_step4}, by sem. of $\bigwedge$, $B_{\pi_k\sigma}$} \label{it:lem_sigmaHC_step5}
			\item $(\intsys, r, t) \models \overline{H}_{\pi_{k+1} \sigma \circ \ldots \circ \pi_{|\sigma|}\sigma} \varphi$ \proofarg{from lines \eqref{it:lem_sigmaHC_step2}, \eqref{it:lem_sigmaHC_step5}, by sem. of $B_{\pi_k\sigma}$, $\bigwedge$, reflexivity of $\pwrelation{\pi_k\sigma}$} \label{it:lem_sigmaHC_step6}
			\item contradiction! \proofarg{from lines \eqref{it:lem_sigmaHC_step3}, \eqref{it:lem_sigmaHC_step6}} \label{it:lem_sigmaHC_step7}
		\end{compactenum}
\end{proof}

\begin{lemmarep} \label{lem:shortCh}
	For persistent formula $\varphi$, $\chi \in \extension[B]$, $r \in \system{\chi}$, $t \in \mathbb{N}$, agent sequence $\sigma = \sigma_s \circ (i) \circ \sigma_p \in \agseq$ where $\sigma_p \ne \epsilon$,
	\begin{equation}
		\left((\intsys, r, t) \models \bigwedge\limits_{j \in \sigma_s \circ (i)} \correct{j} \text{ and } \sigma \in \recvphiall{\varphi}{r_i(t)}{i} \right) \ \Rightarrow \ \sigma_p \in \recvphiall{\varphi}{r_i(t)}{i}
	\end{equation}
\end{lemmarep}
\begin{proof}
	by contradiction.
	\begin{compactenum}
		\item $\sigma_p \ne \epsilon$ \proofarg{contr. assumption} \label{it:lem_shortCh1a}
		\item $\sigma_s \circ (i) \circ \sigma_p \in \recvphiall{\varphi}{r_i(t)}{i}$ \proofarg{by contr. assumption} \label{it:lem_shortCh1}
		\item $(\forall j \in \sigma_s \circ (i))\ (\intsys, r, t) \models \correct{j}$ \proofarg{by contr. assumption} \label{it:lem_shortCh2}
		\item $\sigma_p \not\in \recvphiall{\varphi}{r_i(t)}{i}$ \proofarg{by contr. assumption} \label{it:lem_shortCh3}
		\item $(\intsys, r, t) \models \overline{H}_{(i) \circ \sigma_p} \varphi$ \proofarg{from lines \eqref{it:lem_shortCh1}, \eqref{it:lem_shortCh2}, by Lemma \ref{lem:sigmaHC} for $k = |\sigma_s| + 1$, since in Lemma \ref{lem:sigmaHC} the hope chain is extended by singleton sequence $(i)$} \label{it:lem_shortCh4}
		\item $(\intsys, r, t) \models \trueoccurred[i]{o} \text{ and } \left( (\intsys, r, t) \models \trueoccurred[i]{o} \Rightarrow \overline{H}_{(i) \circ \sigma_p} \varphi \right)$ \proofarg{from line \eqref{it:lem_shortCh4}, by \cite[Theorem 13]{KPSF19:TARK}, by exist. inst. for $o$} \label{it:lem_shortCh5}
	\item $o = \recv{\pi_1\sigma_p}{\mu} \text{ and } \mu \in \msgsphi{\overline{H}_{\pi_2\sigma_p \circ \ldots \circ \pi_{|\sigma_p|}\sigma_p}\varphi}{\pi_1\sigma_p}{i}$ \proofarg{from lines \eqref{it:lem_shortCh1a}, \eqref{it:lem_shortCh5}\footnote{Note that even if the information about $\overline{H}_{(i) \circ \sigma_p} \varphi$ propagated to $i$ via a local edge in the reliable causal cone \cite{KPSF19:TARK} in $\sigma$, this information must have reached agent $i$ initially via some message if $\sigma_p \ne \epsilon$.}, by Definition \ref{def:phi_msgs}, by Definition \ref{def:phi_msgs}, sem. of $H$, $B$} \label{it:lem_shortCh6}
		\item $\recv{\pi_1\sigma_p}{\mu} \in r_i(t)$ \proofarg{from lines \eqref{it:lem_shortCh5}, \eqref{it:lem_shortCh6}, by sem. of $\trueoccurred[i]{o}$} \label{it:lem_shortCh7}
		\item $\sigma_p \in \recvphiall{\varphi}{r_i(t)}{i}$ \proofarg{from lines \eqref{it:lem_shortCh6}, \eqref{it:lem_shortCh7}, by Definition \ref{def:recv_phi_all}} \label{it:lem_shortCh8}
		\item contradiction! \proofarg{from lines \eqref{it:lem_shortCh3}, \eqref{it:lem_shortCh8}} \label{it:lem_shortCh9}
	\end{compactenum}
\end{proof}

Since by Lemma \ref{lem:shortCh} it is sufficient to consider only hope chains where no agent appears twice, we define the appropriate subset of the one in Definition \ref{def:recv_phi_all}.
\begin{definition}  \label{def:recv_phi}
	\begin{equation}
		\begin{aligned}
			\recvphi{\varphi}{r_i(t)}{i} \ \ce \ \{\sigma \in \recvphiall{\varphi}{r_i(t)}{i} \mid\ (\forall k,k'\neq k \in [1, |\sigma|])\ \pi_k \sigma \ne \pi_{k'} \sigma\}
		\end{aligned}
	\end{equation}
\end{definition}

Since this refined set is a subset from \eqref{eq:recv_phi_all}, we immediately get the following result.
\begin{corollary} \label{cor:recv_phi_BH}
	For $\chi \in \extension[B]$, $\intsys = (\system{\chi}, \pi)$, $r \in \system{\chi}$, $t \in \mathbb{N}$, $i,j \in \agents$, and persistent $\varphi$,
	\begin{equation} \label{eq:cor_recv_phi_BH}
		(\forall \sigma \in \recvphi{\varphi}{r_i(t)}{i})\ (\intsys, r, t) \models B_i\overline{H}_\sigma \varphi.
	\end{equation}
\end{corollary}

Clearly, $\recvphi{\varphi}{r_i(t)}{i}$ can be easily computed by agent $i$ in finite time (as opposed to $\recvphiall{\varphi}{r_i(t)}{i}$), which may contain an infinite number of hope chains), 
by checking its local history for the appropriate message receptions for all agents involved
in some hope chain.

\medskip

Obviously, if just one agent involved in a hope chain is faulty, the receiving 
agent $i$ cannot infer anything meaningful from it. In a byzantine asynchronous agent context
$\chi \in \extension[B_f]$, 
where at most $f$ agents can become faulty, it 
thus makes sense to use multiple hope chains leading to $i$ for implementing \emph{fault-tolerant
communication}. In order to be effective, though, the agents appearing in different hope chains 
must be different.

We overload the set difference operator $\setminus$ for sets of sequences and sets as follows:
\begin{definition} \label{def:seqSetDiff}
	Given a set of sequences $\Sigma = \{\sigma_1, \ldots, \sigma_k\}$, where $\sigma_\ell = (i_{\ell_1}, \ldots, i_{\ell_k})$, and set $S = \{j_1, \ldots, j_m\}$, let
	\begin{equation} \label{eq:seqSetDiff}
		\Sigma \setminus S \ce \{\sigma \in \Sigma \mid (\forall i \in \footnote{By slight abuse of notation we write $i \in \sigma$ if{f} $i$ appears somewhere in the sequence $\sigma$.} \sigma)\ i \notin S\}.
	\end{equation}
\end{definition}

\begin{definition} \label{def:powSetOfMaxDisjSeqSets}
	Given a set of sequences $\Sigma = \{\sigma_1, \ldots, \sigma_k\}$, we define the set of all sets of \emph{disjoint sequences} in $\Sigma$, which are disjoint in the sense that no sequence contains an element of another sequence from the same set, as
	\begin{equation} \label{eq:powSetOfDisjSeqSets}
		\disjss{\Sigma} \ce \{\overline{\Sigma} \subseteq \Sigma \mid\ (\forall \sigma', \sigma''\neq\sigma' \in \overline{\Sigma})(\forall k' \in \{1, \ldots, |\sigma'|\})\ \pi_{k'}\sigma' \not\in \sigma''\}.
	\end{equation}
\end{definition}

The following \cref{thm:recv_phi} shows that fault-tolerant communication
via multiple hope chains is effective for agent $i$ to obtain $B_i \varphi$,
provided there are sufficiently many disjoint ones:

\begin{theoremrep} \label{thm:recv_phi}
For agent context $\chi \in \extension[B_f]$, run $r \in \system{\chi}$, timestamp $t \in \mathbb{N}$, agent $i \in \agents$, persistent formula $\varphi$, and a set of agents $F \subseteq \agents$ who $i$ believes to be faulty,
	\begin{equation} \label{eq:lem_recv_phi}
		(\exists \Sigma' \in \disjss{\recvphi{\varphi}{r_i(t)}{i} \setminus F})\ |\Sigma'| \ > \ f - |F| \quad \Rightarrow \quad (\intsys, r, t) \models B_i \varphi.
	\end{equation}
\end{theoremrep}
\begin{proof}
	\begin{compactenum}
		\item $(\forall k \in F)\ (\intsys, r, t) \models B_i \faulty{k}$ \proofarg{by assumption} \label{it:lem_recv_phi_1}
		\item $(\exists \Sigma' \in \disjss{\recvphi{\varphi}{r_i(t)}{i} \setminus F})\ |\Sigma'| \ > \ f - |F|$ \proofarg{by assumption} \label{it:lem_recv_phi_2}
		\item $(\intsys, r, t) \not\models B_i \varphi$ \proofarg{by assumption} \label{it:lem_recv_phi_3}
		\item $\Sigma \in \disjss{\recvphi{\varphi}{r_i(t)}{i} \setminus F}$ \proofarg{simultaneous with line \eqref{it:lem_recv_phi_7}, from line \eqref{it:lem_recv_phi_2} , by exist. inst.} \label{it:lem_recv_phi_6}
		\item $|\Sigma| \ > \ f - |F|$ \proofarg{simultaneous with line \eqref{it:lem_recv_phi_6}, from line \eqref{it:lem_recv_phi_2}, by exist. inst.} \label{it:lem_recv_phi_7}
		\item $(\forall \sigma' \in \Sigma)\ (\intsys, r, t) \models B_i\overline{H}_{\sigma'} \varphi$ \proofarg{from line \eqref{it:lem_recv_phi_6}, by Corollary \ref{cor:recv_phi_BH} and Definition \ref{def:powSetOfMaxDisjSeqSets}} \label{it:lem_recv_phi_4}
		\item $\widetilde{r} \in \system{\chi} \text{ and } \widetilde{t} \in \mathbb{N} \text{ and } r(t) \pwrelation{i} \widetilde{r}(\widetilde{t}) \text{ and } (\intsys, \widetilde{r}, \widetilde{t}) \models \correct{i} \wedge \neg\varphi$ \proofarg{from line \eqref{it:lem_recv_phi_3}, by sem. of $B_i$, $\not\models$, exist. inst.} \label{it:lem_recv_phi_8}
		\item $(\forall k \in F)\ (\intsys, \widetilde{r}, \widetilde{t}) \models \correct{i} \rightarrow \faulty{k}$ \proofarg{from lines \eqref{it:lem_recv_phi_1}, \eqref{it:lem_recv_phi_8}, by sem. of $B_i$, $\Rightarrow$, $\wedge$, $\pwrelation{i}$, univ. inst.} \label{it:lem_recv_phi_9}
		\item $(\forall \sigma' \in \Sigma)\ (\intsys, \widetilde{r}, \widetilde{t}) \models \correct{i} \rightarrow \overline{H}_{\sigma'} \varphi$ \proofarg{from lines \eqref{it:lem_recv_phi_4}, \eqref{it:lem_recv_phi_8}, by definition of $\pwrelation{i}$, sem. of $B_i$, $\Rightarrow$, $\wedge$, univ. inst.} \label{it:lem_recv_phi_10}
	\item $(\forall k \in F)\ (\intsys, \widetilde{r}, \widetilde{t}) \models \faulty{k}$ \proofarg{from lines \eqref{it:lem_recv_phi_8}, \eqref{it:lem_recv_phi_9}, by sem. of $\rightarrow$, ``and'', $\wedge$} \label{it:lem_recv_phi_11}
	\item $(\forall \sigma' \in \Sigma)\ (\intsys, \widetilde{r}, \widetilde{t}) \models \overline{H}_{\sigma'} \varphi$ \proofarg{from lines \eqref{it:lem_recv_phi_8}, \eqref{it:lem_recv_phi_10}, by sem. of $\rightarrow$, ``and'', $\wedge$} \label{it:lem_recv_phi_12}
	\item $\sigma \in \Sigma \text{ and } (\forall l \in \{1, \ldots, |\sigma|\})\ (\intsys, \widetilde{r}, \widetilde{t}) \models \correct{\pi_l\sigma}$ \proofarg{\eqref{it:lem_recv_phi_6}, \eqref{it:lem_recv_phi_7}, \eqref{it:lem_recv_phi_8}, \eqref{it:lem_recv_phi_11}, by def. of $\correct{k}$, $\faulty{k}$, $\filtere[B_f]{}{}$, $\forall$, $\pwrelation{i}$, exist. inst., via pigeonhole argument} \label{it:lem_recv_phi_13}
	\item $(\intsys, \widetilde{r}, \widetilde{t}) \models \varphi$ \proofarg{from lines \eqref{it:lem_recv_phi_12}, \eqref{it:lem_recv_phi_13}, by sem. of ``and'', $\forall$, $\rightarrow$, $H$, Definition \ref{def:hopeChain}, univ. inst. and reflexivity of $\pwrelation{\widetilde{j}}$} \label{it:lem_recv_phi_14}
	\item contradiction! \proofarg{from lines \eqref{it:lem_recv_phi_8}, \eqref{it:lem_recv_phi_14}, by sem. of ``and'', $\wedge$} \label{it:lem_recv_phi_15}
	\end{compactenum}
\end{proof}

Apart from being informed about $\varphi$ via fault-tolerant communication, agent $i$ can of
course also obtain the belief\footnote{Be aware of the following validities: $K_i \varphi \ \Rightarrow \ B_i \varphi \quad \text{and} \quad K_i \occurred[i]{o} \ \Rightarrow \ B_i \trueoccurred[i]{o}$ (following directly from the definition of the belief modality).} $B_i\varphi$ by observing $\varphi$ directly in its local history, i.e., when
$\varphi$ is local at $i$:

\begin{theoremrep} \label{thm:loc_cond_bel_phi}
	For agent context $\chi \in \extension[B]$, interpreted system $\intsys = (\system{\chi}, \pi)$, run $r \in \system{\chi}$, timestamp $t \in \mathbb{N}$, agent $i \in \agents$, local state $\lambda_0 \in \localinitialstates[i]$, and action or event $o \in \haps[i]$,
	\begin{equation} \label{eq:thm_loc_cond_bel_phi}
		\begin{aligned}
			o \in r_i(t) \quad &\Rightarrow \quad (\intsys, r, t) \models K_i \occurred[i]{o} \\
			\lambda_0 = r_i(0) \quad &\Rightarrow \quad (\intsys, r, t) \models K_i \init{i}{\lambda_0}
		\end{aligned}
	\end{equation}
\end{theoremrep}
\begin{proof}
	Regarding the first line:
	\begin{compactenum}
		\item $o \in r_i(t)$ \proofarg{by contr. assumption} \label{proof_first_line_thm_loc_cond_bel_phi_it1}
		\item $(\intsys, r, t) \not\models K_i \occurred[i]{o}$ \proofarg{by contr. assumption} \label{proof_first_line_thm_loc_cond_bel_phi_it2}
		\item $\widetilde{r} \in \system{\chi} \text{ and } \widetilde{t} \in \mathbb{N} \text{ and } r(t) \pwrelation{i} \widetilde{r}(\widetilde{t}) \text{ and } (\intsys, \widetilde{r}, \widetilde{t}) \not\models \occurred[i]{o}$ \proofarg{from line \eqref{proof_first_line_thm_loc_cond_bel_phi_it2}, by sem. of $\not\models$, $K_i$, exist. inst.} \label{proof_first_line_thm_loc_cond_bel_phi_it3}
		\item $r_i(t) = \widetilde{r}_i(\widetilde{t})$ \proofarg{from line \eqref{proof_first_line_thm_loc_cond_bel_phi_it3}, by sem. of $\pwrelation{i}$} \label{proof_first_line_thm_loc_cond_bel_phi_it4}
		\item $o \in \widetilde{r}(\widetilde{t})$ \proofarg{from lines \eqref{proof_first_line_thm_loc_cond_bel_phi_it1}, \eqref{proof_first_line_thm_loc_cond_bel_phi_it4}, by sem. of $=$} \label{proof_first_line_thm_loc_cond_bel_phi_it5}
		\item $(\intsys, \widetilde{r}, \widetilde{t}) \models \occurred[i]{o}$ \proofarg{from line \eqref{proof_first_line_thm_loc_cond_bel_phi_it5}, by sem. of $\occurred[i]{o}$} \label{proof_first_line_thm_loc_cond_bel_phi_it6}
		\item contradiction! \proofarg{from lines \eqref{proof_first_line_thm_loc_cond_bel_phi_it3}, \eqref{proof_first_line_thm_loc_cond_bel_phi_it6}} \label{proof_first_line_thm_loc_cond_bel_phi_it7}
	\end{compactenum}

	Regarding the second line:
	\begin{compactenum}
		\item $\lambda_0 = r_i(0)$ \proofarg{by contr. assumption} \label{proof_second_line_thm_loc_cond_bel_phi_it1}
		\item $(\intsys, r, t) \not\models K_i \init{i}{\lambda_0}$ \proofarg{by contr. assumption} \label{proof_second_line_thm_loc_cond_bel_phi_it2}
		\item $\widetilde{r} \in \system{\chi} \text{ and } \widetilde{t} \in \mathbb{N} \text{ and } r(t) \pwrelation{i} \widetilde{r}(\widetilde{t}) \text{ and } (\intsys, \widetilde{r}, \widetilde{t}) \not\models \init{i}{\lambda_0}$ \proofarg{from line \eqref{proof_second_line_thm_loc_cond_bel_phi_it2}, by sem. of $\not\models$, $K_i$, exist. inst.} \label{proof_second_line_thm_loc_cond_bel_phi_it3}
		\item $r_i(t) = \widetilde{r}_i(\widetilde{t})$ \proofarg{from line \eqref{proof_second_line_thm_loc_cond_bel_phi_it3}, by sem. of $\pwrelation{i}$} \label{proof_second_line_thm_loc_cond_bel_phi_it4}
		\item $\lambda_0 = \widetilde{r}(\widetilde{0})$ \proofarg{from lines \eqref{proof_second_line_thm_loc_cond_bel_phi_it1}, \eqref{proof_second_line_thm_loc_cond_bel_phi_it4}, by sem. of $=$} \label{proof_second_line_thm_loc_cond_bel_phi_it5}
		\item $(\intsys, \widetilde{r}, \widetilde{t}) \models \init{i}{\lambda_0}$ \proofarg{from line \eqref{proof_second_line_thm_loc_cond_bel_phi_it5}, by sem. of $\init{i}{\lambda_0}$} \label{proof_second_line_thm_loc_cond_bel_phi_it6}
		\item contradiction! \proofarg{from lines \eqref{proof_second_line_thm_loc_cond_bel_phi_it3}, \eqref{proof_second_line_thm_loc_cond_bel_phi_it6}} \label{proof_second_line_thm_loc_cond_bel_phi_it7}
	\end{compactenum}
\end{proof}

\section{Belief Gain About Faultiness}
\label{sec:faults}

In this section, we will address the question of how agent $i$ can establish 
belief about some agent $j$ being faulty. In line with \cref{thm:recv_phi} and \cref{thm:loc_cond_bel_phi},
there are two ways of achieving this: by direct observation, namely, receiving an obviously faulty 
message from $j$, or by receiving trustworthy notifications about $j$'s faultiness
from sufficiently many other agents. We start with the former case.

\begin{lemmarep}[Directly observing others' faults] \label{lem:cond_belief_others_faulty}
	For interpreted system $\intsys = (\system{\chi}, \pi)$ with agent context 
$\chi = ((P_\epsilon, \globalinitialstates, \tauprotocol{B}{\envprotocol{}}{\joinprotocol{}}, \Admissibility{}), P)$, 
run $r \in \system{\chi}$, timestamp $t \in \mathbb{N}$ and agents $i, j \in \agents$,  
if $(\exists \mu \in \Msgs)(\forall h_{j} \in \localstates{j})(\forall D \in P_{j}(h_{j}))
	\send{i}{\mu} \notin D \ \wedge \ \recv{j}{\mu} \in r_i(t)$, then $(\intsys, r, t) \models B_i \faulty{j}$.
\end{lemmarep}
\begin{proof}
	\begin{compactenum}
		\item $(\exists \mu \in \Msgs)(\forall h_{j} \in \localstates{j})(\forall D \in P_{j}(h_{j}))\ \send{i}{\mu} \notin D$ \proofarg{by contr. assumption} \label{it:lem_cond_belief_others_faulty_1}
		\item $\recv{j}{\mu} \in r_i(t)$ \proofarg{by contr. assumption} \label{it:lem_cond_belief_others_faulty_2}
		\item $(\intsys, r, t) \not\models B_i \faulty{j}$ \proofarg{by contr. assumption} \label{it:lem_cond_belief_others_faulty_3}
		\item $(\forall \widetilde{r} \in \system{\chi})(\forall \widetilde{t} \in \mathbb{N})\ (\intsys, \widetilde{r}, \widetilde{t}) \not\models \trueoccurred[j]{\send{i}{\mu}}$ \proofarg{from line \eqref{it:lem_cond_belief_others_faulty_1}, by def. of $\trueoccurred{j}$} \label{it:lem_cond_belief_others_faulty_4}
		\item $\widehat{r} \in \system{\chi} \text{ and } \widehat{t} \in \mathbb{N} \text{ and } r(t) \pwrelation{i} \widehat{r}(\widehat{t}) \text{ and }$ \\
			$(\intsys, \widehat{r}, \widehat{t}) \models \correct{i} \ \wedge \ \correct{j}$ \proofarg{from line \eqref{it:lem_cond_belief_others_faulty_3}, by sem. of $B_i$, $\not\models$ and exist. inst.} \label{it:lem_cond_belief_others_faulty_5}
		\item $r_i(t) = \widehat{r}_i(\widehat{t})$ \proofarg{from line \eqref{it:lem_cond_belief_others_faulty_5}, by def. of $\pwrelation{i}$} \label{it:lem_cond_belief_others_faulty_6}
		\item $(\intsys, \widehat{r}, \widehat{t}) \models \correct{i}$ \proofarg{from line \eqref{it:lem_cond_belief_others_faulty_5}, by sem. of $\wedge$} \label{it:lem_cond_belief_others_faulty_7}
		\item $(\intsys, \widehat{r}, \widehat{t}) \models \correct{j}$ \proofarg{from line \eqref{it:lem_cond_belief_others_faulty_5}, by sem. of $\wedge$} \label{it:lem_cond_belief_others_faulty_8}
		\item $(\intsys, r, t) \models \occurred[i]{\recv{j}{\mu}}$ \proofarg{from line \eqref{it:lem_cond_belief_others_faulty_2}, by def. of $\occurred[i]{}$} \label{it:lem_cond_belief_others_faulty_9}
		\item $(\intsys, \widehat{r}, \widehat{t}) \models \occurred[i]{\recv{j}{\mu}}$ \proofarg{from line \eqref{it:lem_cond_belief_others_faulty_6}, \eqref{it:lem_cond_belief_others_faulty_9}, by sem. of $=$, $\occurred[i]{}$} \label{it:lem_cond_belief_others_faulty_10}
		\item $(\intsys, \widehat{r}, \widehat{t}) \models \trueoccurred[i]{\recv{j}{\mu}} \ \vee \ \fake[i]{\recv{j}{\mu}}$ \proofarg{from line \eqref{it:lem_cond_belief_others_faulty_10}, by def. of $\occurred[i]{}$} \label{it:lem_cond_belief_others_faulty_11}
		\item
		\begin{compactenum}
			\item $(\intsys, \widehat{r}, \widehat{t}) \models \trueoccurred[i]{\recv{j}{\mu}}$ \proofarg{from line \eqref{it:lem_cond_belief_others_faulty_11}, by sem. of $\vee$} \label{it:lem_cond_belief_others_faulty_12.1}
			\item $(\intsys, \widehat{r}, \widehat{t}) \models \fhappened[j]{\send{i}{\mu}}$ \proofarg{from lines \eqref{it:lem_cond_belief_others_faulty_4}, \eqref{it:lem_cond_belief_others_faulty_12.1} by def. of $\filtere[B]{}{}$} \label{it:lem_cond_belief_others_faulty_12.2}
			\item $(\intsys, \widehat{r}, \widehat{t}) \models \faulty{j}$ \proofarg{from line \eqref{it:lem_cond_belief_others_faulty_12.2}, by def. of $\fhappened[j]{}$} \label{it:lem_cond_belief_others_faulty_12.3}
			\item contradiction! \proofarg{from lines \eqref{it:lem_cond_belief_others_faulty_8}, \eqref{it:lem_cond_belief_others_faulty_12.3}} \label{it:lem_cond_belief_others_faulty_12.4}
		\end{compactenum}
		\item
		\begin{compactenum}
			\item $(\intsys, \widehat{r}, \widehat{t}) \models \fake[i]{\recv{j}{\mu}}$ \proofarg{from line \eqref{it:lem_cond_belief_others_faulty_11}, by sem. of $\vee$} \label{it:lem_cond_belief_others_faulty_13.1}
			\item $(\intsys, \widehat{r}, \widehat{t}) \models \faulty{i}$ \proofarg{from line \eqref{it:lem_cond_belief_others_faulty_13.1}, by def. of $\fake[i]{}$} \label{it:lem_cond_belief_others_faulty_13.2}
			\item contradiction! \proofarg{from lines \eqref{it:lem_cond_belief_others_faulty_7}, \eqref{it:lem_cond_belief_others_faulty_13.2}} \label{it:lem_cond_belief_others_faulty_13.3}
		\end{compactenum}
	\end{compactenum}
\end{proof}

Note carefully that the messages that allow direct fault detection 
in \cref{lem:cond_belief_others_faulty} must indeed be obviously
faulty, in the sense that they must not occur in \emph{any} correct run.
This is the case for messages that report some local history of the sending
agent that is inconsistent with the local history communicated
earlier, which covers the fault detection requirements in
\cite{MTH14:STOC}, for example. However, messages that could 
not occur just in the \emph{specific} run $r\in \system{\chi}$ cannot
be used for direct fault detection. We capture this by the following 
characterization of directly detectable faults: 

\begin{definition} \label{def:DirObBelFaultyAg}
	For some agent context $\chi = ((P_\epsilon, \globalinitialstates, \tauprotocol{B}{\envprotocol{}}{\joinprotocol{}}, \Admissibility{}), P) \in \extension[B]$, agent $i$ and local history $h_i \in \localstates{i}$, the set of agents that $i$ beliefs to be faulty, 
due to having received an obviously faulty message from them, is defined as
	\begin{equation} \label{eq:DirObBelFaultyAg}
		\begin{aligned}
			\dirobfag(h_i, i) \ce \{j \in \agents \mid &(\exists \mu \in \msgs)(\forall h_j \in \localstates{j})(\forall D \in P_j(h_j)) \\
													   &\send{i}{\mu} \notin D \wedge \recv{j}{\mu} \in h_i\}.
		\end{aligned}
	\end{equation}
\end{definition}

\begin{corollary} \label{cor:DirObBelFaultyAg_belief}
	For some $\chi = ((P_\epsilon, \globalinitialstates, \tauprotocol{B}{\envprotocol{}}{\joinprotocol{}}, \Admissibility), P) \in \extension[B]$, $\I = (\system{\chi}, \pi)$, $r \in \system{\chi}$, $t \in \mathbb{N}$, $i \in \agents$, if
	\begin{equation} \label{eq:DirObBelFaultyAg_belief}
		j \in \dirobfag(r_i(t), i) \ \Rightarrow \ (\I, r, t) \models B_i \faulty{j}.
	\end{equation}
\end{corollary}


Turning our attention to the case corresponding to Theorem \ref{thm:recv_phi}, namely, detecting faultiness of
an agent by receiving sufficiently many trustworthy notifications from other agents in an agent context $\chi \in \extension[B_f]$, it seems obvious to
use the following characterization:
\begin{equation} \label{eq:NotifBelFaultyAg}
	\begin{aligned}
		\notifbfag(h_i, i) \ce\ \{\ell \in \agents \mid (\exists \Sigma \in \disjss{\recvphi{\faulty{\ell}}{h_i}{i} \setminus \notifbfag(h_i, i)})\ |\Sigma| \ > \ f - |\notifbfag(h_i, i)| \}.
	\end{aligned}
\end{equation}
Indeed, if there are more than $f$ minus the currently believed faulty agents disjoint hope chains for $\faulty{\ell}$
leading to agent $i$, it can safely add agent $\ell$ to its set of currently believed faulty agents. Unfortunately, however,
this definition is cyclic, as $\notifbfag( h_i, i)$ appears in its own definition (with the exact same parameters):
Who an agent believes to be faulty depends on who an agent already believes to be faulty. We will get rid of this problem
by a fixpoint formulation, which can be solved algorithmically.

We start out from direct notifications by a faulty agent, i.e., when an agent $j$ that has
somehow detected its own faultiness (see below) informs agent $i$ about this fact.
Note that this is different from \cref{cor:DirObBelFaultyAg_belief}, where agent $i$ directly
observes $j$'s misbehavior ($i$ received an obviously faulty message).


We capture this by the following characterization of directly notified faults: 

\begin{definition} \label{def:DirNotifBelFaultyAg}
	For some agent context $\chi \in \extension[B_f]$, agent $i \in \agents$ and local state $h_i \in \localstates{i}$, we define the set of agents, who agent $i$ believes to be faulty due having received a direct notification from exactly those agents, as
	\begin{equation} \label{eq:def_dirnotifbfag}
		\dirnotifbfag(h_i, i) \ce \{j \in \agents \mid (j) \in \recvphi{\faulty{j}}{h_i}{i} \}.
	\end{equation}
\end{definition}

\begin{lemmarep} \label{cor:DirNotifBelFaultyAg}
	For some  agent context $\chi = ((P_\epsilon, \globalinitialstates, \tauprotocol{B}{\envprotocol{}}{\joinprotocol{}}, \Admissibility), P) \in \extension[B]$, $\I = (\system{\chi}, \pi)$, run $r \in \system{\chi}$, $t \in \mathbb{N}$ and  agent $i \in \agents$, if
	\begin{equation} \label{eq:DirNotifBelFaultyAg}
		j \in \dirnotifbfag(r_i(t), i) \ \Rightarrow \ (\I, r, t) \models B_i \faulty{j}.
	\end{equation}
\end{lemmarep}
\begin{proof}
	\begin{compactenum}
		\item $(j) \in \recvphi{\faulty{j}}{r_i(t)}{i}$ \proofarg{by contr. assumption, Definition \ref{def:DirNotifBelFaultyAg}} \label{it:lem_DirNotifBelFaultyAg1}
		\item $(\I, r, t) \not\models B_i \faulty{j}$ \proofarg{by contr. assumption, Definition \ref{def:DirNotifBelFaultyAg}} \label{it:lem_DirNotifBelFaultyAg2}
		\item $\widetilde{r} \in \system{\chi} \text{ and } \widetilde{t} \in \mathbb{N} \text{ and } r(t) \pwrelation{i} \widetilde{r}(\widetilde{t})$ \proofarg{from line \eqref{it:lem_DirNotifBelFaultyAg2}, by sem. of $B_i$ and exist. inst.} \label{it:lem_DirNotifBelFaultyAg3}
		\item $(\I, \widetilde{r}, \widetilde{t}) \models \correct{i} \wedge \correct{j}$ \proofarg{from line \eqref{it:lem_DirNotifBelFaultyAg2}, by sem. of $B_i$ and exist. inst.} \label{it:lem_DirNotifBelFaultyAg4}
		\item $\mu \in \msgsphi{\faulty{j}}{j}{i}$ \proofarg{from line \eqref{it:lem_DirNotifBelFaultyAg1}, by Definition \ref{def:recv_phi} and exist. inst. for $\mu$} \label{it:lem_DirNotifBelFaultyAg5}
		\item $\recv{j}{\mu} \in r_i(t)$ \proofarg{from line \eqref{it:lem_DirNotifBelFaultyAg1}, by Definition \ref{def:recv_phi} and exist. inst. for $\mu$} \label{it:lem_DirNotifBelFaultyAg6}
		\item $\recv{j}{\mu} \in \widetilde{r}_i(\widetilde{t})$ \proofarg{from lines \eqref{it:lem_DirNotifBelFaultyAg3}, \eqref{it:lem_DirNotifBelFaultyAg6}, by def. of $\pwrelation{i}$} \label{it:lem_DirNotifBelFaultyAg7}
		\item $(\I, \widetilde{r}, \widetilde{t}) \models \trueoccurred[i]{\recv{j}{\mu}}$ \proofarg{from lines \eqref{it:lem_DirNotifBelFaultyAg4}, \eqref{it:lem_DirNotifBelFaultyAg7}, by sem. of $\correct{i}$, $\wedge$} \label{it:lem_DirNotifBelFaultyAg8}
		\item $(\I, \widetilde{r}, \widetilde{t}) \models \trueoccurred[j]{\send{i}{\mu}}$ \proofarg{from lines \eqref{it:lem_DirNotifBelFaultyAg4}, \eqref{it:lem_DirNotifBelFaultyAg8}, by def. of $\filtere[B]{}{}$, sem. of $\correct{j}$ and $\wedge$} \label{it:lem_DirNotifBelFaultyAg9}
		\item $(\I, \widetilde{r}, \widetilde{t}) \models B_j \faulty{j}$ \proofarg{from lines \eqref{it:lem_DirNotifBelFaultyAg5}, \eqref{it:lem_DirNotifBelFaultyAg9}, by Definition \ref{def:phi_msgs}} \label{it:lem_DirNotifBelFaultyAg10}
		\item $(\I, \widetilde{r}, \widetilde{t}) \models \faulty{j}$ \proofarg{from lines \eqref{it:lem_DirNotifBelFaultyAg4}, \eqref{it:lem_DirNotifBelFaultyAg10}, by sem. of $B_j$, $\wedge$, $\rightarrow$, reflexivity of $\pwrelation{j}$} \label{it:lem_DirNotifBelFaultyAg11}
		\item contradiction! \proofarg{from lines \eqref{it:lem_DirNotifBelFaultyAg4}, \eqref{it:lem_DirNotifBelFaultyAg11}, sem. of $\wedge$} \label{it:lem_DirNotifBelFaultyAg12}
	\end{compactenum}
\end{proof}

As it is possible for an agent $i$ in an agent context $\chi \in \extension[B_f]$
to sometimes also detect its own faultiness \cite{KPSF19:FroCos}, we need to consider
this as well. The following function returns $\true$ if the agent $i$ observes some
erroneous behavior in its own history $h_i=(h_i(|h_i|),h_i(|h_i|-1),\dots,h_i(0))$
(recall that $h_i(k)$ is the set of all haps agent $i$ perceived in the $k$-th 
round it was active in), which implies that $i$ itself is faulty:

\begin{definition} \label{def:DirObMeKnowFaulty}
For any $\chi \in \extension$, agent $i \in \agents$, timestamp $t \in \mathbb{N}$, and local history $h_i=r_i(t)$, let
	\begin{equation}
		\dirobmkf(r_i(t), i) \ce
		\begin{cases}
			\true \quad \text{if } (\exists a \in \actions[i])(\exists m \in [1, |h_i|-1]) \\
			\phantom{\true \quad \text{if }}(\forall D \in P_i(\pi_m h_i, \pi_{m+1} h_i, \ldots, \pi_{|h_i|} h_i)\\
			\phantom{\true \quad \text{if }}a \notin D \ \wedge \ a \in \pi_{m+1} h_i \\
			\false \quad \text{otherwise}.
		\end{cases}
	\end{equation}
\end{definition}

\begin{corollary}[Observing one's own faulty history] \label{cor:MeKnowFaulty}
	For
	an agent context $\chi \in \extension[B]$, interpreted system $\I = (\system{\chi}, \pi)$,
	run $r \in \system{\chi}$, timestamp $t \in \mathbb{N}$, and agent $i \in \agents$,
	 if
	\begin{equation} \label{eq:MeKnowFaulty}
		\dirobmkf(r_i(t), i) = \true \quad \Rightarrow \quad (\I, r, t) \models K_i \faulty{i}
	\end{equation}
\end{corollary}

With these preparations, we are now ready to present a procedure, given in 
in Algorithm~\ref{alg:GBAlgFaultyAgs}, by which any agent $i$ can compute 
its belief regarding the faultiness of agents (including itself). Rather than
explicitly constructing the underlying Kripke model, it exploits the \emph{a
priori} knowledge of the sets resp.\ the function in Definition \ref{def:phi_msgs}, \ref{def:DirObBelFaultyAg}, \ref{def:DirNotifBelFaultyAg} resp.\ \ref{def:DirObMeKnowFaulty}. Note carefully
that they can indeed be pre-computed ``offline' and supplied to the algorithm
via the resulting look-up tables. In sharp contrast to constructing the Kripke
model, our procedure is guaranteed to terminate in a bounded number of steps.

\begin{savenotes}
\begin{algorithm}
	\caption{Gain-belief algorithm for faulty agents in agent context $\chi \in \extension[B_f]$ for agent $i$ with history $h_i$}
	\label{alg:GBAlgFaultyAgs}
	\begin{algorithmic}[1]
		\Function{BeliefWhoIsFaultyAlgorithm}{$\chi, h_i, i, f$}
			\State $F \ce \dirobfag(h_i, i) \cup \dirnotifbfag(h_i, i)$
			\If{$\dirobmkf(h_i, i) = true$}
				\State $F \ce F \cup \{i\}$
			\EndIf
			\Repeat
				\State $F_{Old} \ce F$
				\For{all $\ell \in \agents \setminus F$}
					\For{all $\Sigma \in \disjss{\recvphi{\faulty{\ell}}{h_i}{i} \setminus F}$}
						\If{$|\Sigma| > f - |F|$}
							\State $F \ce F \cup \{\ell\}$
							\State \textbf{continue} next iteration at line 7
						\EndIf
					\EndFor
				\EndFor
			\Until{$F = F_{Old}$}
			\State \textbf{return} F
		\EndFunction
	\end{algorithmic}
\end{algorithm}
\end{savenotes}

We start our correctness proof of Algorithm~\ref{alg:GBAlgFaultyAgs} by showing the following invariant
of the set $F$:

\begin{lemmarep} \label{lem:AlgBelH2}
	For Algorithm \ref{alg:GBAlgFaultyAgs} called with parameters $(\chi, h_i, i, f)$, where $\chi \in \extension[B_f]$, interpreted system $\I = (\system{\chi}, \pi)$, $r \in \system{\chi}$, $t \in \mathbb{N}$, $i \in \agents$, and $h_i = r_i(t)$, the following invariant holds for the variable $F$ during its iterations:
	\begin{equation}
		(\forall r \in \system{\chi})(\forall t \in \mathbb{N})(\forall \ell \in F)\ (r_i(t) = h_i) \ \Rightarrow \ (\I, r, t) \models B_i \faulty{\ell}.
	\end{equation}
\end{lemmarep}
\begin{proof}
	By induction over the size of set $F$, $l = |F|$.
	\\
	\underline{Ind. Hyp.:} $(\forall r \in \system{\chi})(\forall t \in \mathbb{N})(\forall \ell \in F)\ (r_i(t) = h_i) \ \Rightarrow \ ((\system{\chi}, \pi), r, t) \models B_i \faulty{\ell}$
	\\
	\underline{Base case for $l = |\dirobfag(h_i, i) \cup \dirnotifbfag(h_i, i)|$:} \proofarg{by code line 2}
	\\
	The induction hypothesis follows from Corollary \ref{cor:DirObBelFaultyAg_belief} and \ref{cor:DirNotifBelFaultyAg}.
	If the condition on code line 3 is true, then the statement additionally follows from Corollary \ref{cor:MeKnowFaulty}.
	\\
	\underline{Ind. Step:}
	Suppose the induction hypothesis holds for $l = |F|$.
	The only line at which $F$ is modified in the main loop (starting at line 5) is line 10.
	From line 8 and 9 in the code we get that there exists a $\Sigma \in \disjss{\recvphi{\faulty{\ell}}{h_i}{i} \setminus F}$ s.t. $|\Sigma| > f - |F|$ before the execution of line 10.
	Hence the induction hypothesis still remains satisfied by Theorem \ref{thm:recv_phi} after line 10 has been executed.
\end{proof}

\begin{corollary} \label{thm:AlgBelief}
	For agent context $\chi \in \extension[B_f]$, interpreted system $\I = (\system{\chi}, \pi)$,
	run $r \in \system{\chi}$,
	timestamp $t \in \mathbb{N}$, and
	agent $i \in \agents$,	
	\begin{equation} \label{eq:cor_AlgBelief}
		\ell \in \texttt{BeliefWhoIsFaultyAlgorithm}(\chi, r_i(t), i, f) \ \Rightarrow \ (\I, r, t) \models B_i \faulty{\ell}.
	\end{equation}
\end{corollary}

Since it follows immediately from the definition of $\extension[B_f]$ that the number of faulty agents in any
run $r \in \system{\chi} \in \extension[B_f]$ is at most $f$, the result of Algorithm~\ref{alg:GBAlgFaultyAgs}
respects $f$ as well:

\begin{lemmarep} \label{lem:corr_ag_bel_faulty_le_f}
	For $\chi \in \extension[B_f]$, $r \in \system{\chi}$, $t \in \mathbb{N}$, correct agent $i \in \agents$, and the set $F$ returned by Algorithm \ref{alg:GBAlgFaultyAgs}\\
	\texttt{BeliefWhoIsFaultyAlgorithm}($\chi, h_i, i, f$), it holds that $|F| \le f$.
\end{lemmarep}
\begin{proof}
	\begin{compactenum}
		\item $|F| > f$ \proofarg{by contr. assumption} \label{it:lem_corr_ag_bel_faulty_le_f_1}
		\item $F =$ \texttt{BeliefWhoIsFaultyAlgorithm}($\chi, h_i, i, f$) \proofarg{by contr. assumption} \label{it:lem_corr_ag_bel_faulty_le_f_2}
		\item $\chi \in \extension[B_f]$ and $r \in \system{\chi}$ and $t \in \mathbb{N}$ \proofarg{by contr. assumption} \label{it:lem_corr_ag_bel_faulty_le_f_3}
		\item $(\intsys, r, t) \models \correct{i}$ \proofarg{by contr. assumption} \label{it:lem_corr_ag_bel_faulty_le_f_4}
		\item $(\forall \ell \in F)\ (\intsys, r, t) \models B_i \faulty{\ell}$ \proofarg{from lines \eqref{it:lem_corr_ag_bel_faulty_le_f_2}, \eqref{it:lem_corr_ag_bel_faulty_le_f_3}, by Theorem \ref{thm:AlgBelief}} \label{it:lem_corr_ag_bel_faulty_le_f_5}
		\item \#faulty agents in $r$ is at most $f$ \proofarg{from line \eqref{it:lem_corr_ag_bel_faulty_le_f_3}, by definition of $\extension[B_f]$} \label{it:lem_corr_ag_bel_faulty_le_f_6}
		\item $j \in F$ and $(\intsys, r, t) \not\models \faulty{j}$ \proofarg{from lines \eqref{it:lem_corr_ag_bel_faulty_le_f_1}, \eqref{it:lem_corr_ag_bel_faulty_le_f_6}, by sem. of $|...|$, $>$, exist. inst.} \label{it:lem_corr_ag_bel_faulty_le_f_7}
		\item $(\intsys, r, t) \models B_i \faulty{j}$ \proofarg{from lines \eqref{it:lem_corr_ag_bel_faulty_le_f_5}, \eqref{it:lem_corr_ag_bel_faulty_le_f_7}, by univ. inst.} \label{it:lem_corr_ag_bel_faulty_le_f_8}
		\item $(\forall r' \in \system{\chi})(\forall t' \in \mathbb{N})\ r(t) \pwrelation{i} r'(t') \ \Rightarrow \ (\intsys, r', t') \models \correct{i} \rightarrow \faulty{j}$ \proofarg{from line \eqref{it:lem_corr_ag_bel_faulty_le_f_8}, by sem. of $B_i$} \label{it:lem_corr_ag_bel_faulty_le_f_9}
		\item $r(t) \pwrelation{i} r(t)$ \proofarg{from line \eqref{it:lem_corr_ag_bel_faulty_le_f_3}, by reflexivity of $\pwrelation{i}$} \label{it:lem_corr_ag_bel_faulty_le_f_10}
		\item $(\intsys, r, t) \models \faulty{j}$ \proofarg{from lines \eqref{it:lem_corr_ag_bel_faulty_le_f_4}, \eqref{it:lem_corr_ag_bel_faulty_le_f_9}, \eqref{it:lem_corr_ag_bel_faulty_le_f_10}, by sem. of $\Rightarrow$, $\rightarrow$ univ. inst.} \label{it:lem_corr_ag_bel_faulty_le_f_11}
		\item contradiction! \proofarg{from lines \eqref{it:lem_corr_ag_bel_faulty_le_f_7}, \eqref{it:lem_corr_ag_bel_faulty_le_f_11}} \label{it:lem_corr_ag_bel_faulty_le_f_12}
	\end{compactenum}
\end{proof}

The following theorem finally proves that Algorithm~\ref{alg:GBAlgFaultyAgs} terminates after a bounded number of steps,
provided the agent context $\chi \in \extension[B_f]$ ensures that agent $i$'s history is finite at every point 
in time, meaning $(\forall t \in \mathbb{N})(\exists b \in \mathbb{N})(\forall t': 0 < t' \le t)\ |r_i(t')| < b$.

\begin{theoremrep} \label{thm:AlgBelTerm}
	For agent context $\chi \in \extension[B_f]$, $\I = (\system{\chi}, \pi)$,
	run $r \in \system{\chi}$,
	timestamp $t \in \mathbb{N}$, and
	agent $i \in \agents$,
	if $\agents$ is finite and $i$'s history is finite at every point in time,  
	then the call $\texttt{BeliefWhoIsFaultyAlgorithm}$\\
	$(\chi, r_i(t), i, f)$ invoking Algorithm \ref{alg:GBAlgFaultyAgs} terminates after a bounded number of steps.
\end{theoremrep}
\begin{proof}
	The fact that variable $F$ in Algorithm \ref{alg:GBAlgFaultyAgs} is monotonically increasing follows from the code lines 2, 4 and 10 as these are the only lines that modify $F$.
	Hence since the set of agents is finite, the outer loop spanning across lines 5. - 12. cannot run forever, as it is bounded by $|\agents|$, since we only add agents from the set $\agents \setminus F$ and in the worst case only one new agent is added per iteration.
	Thus at the latest the loop must terminate, when finally $F = \agents$.
	If during some iteration $F$ doesn't change, the loop terminates early.

	The same argument goes for the loop spanning across lines 7. - 11.

	Regarding the loop spanning lines 8. - 11., since we assumed that agent $i$'s local history is bounded for every local timestamp by some $b \in \mathbb{N}$, agent $i$ could at the most have received messages about $b$ different agent sequences (message chains).
	Since, by Definition \ref{def:powSetOfMaxDisjSeqSets}, $\disjss{\recvphi{\faulty{\ell}}{h_i}{i} \setminus F}$ is a subset of the power set of $\recvphi{\faulty{\ell}}{h_i}{i} \setminus F$ this loop is thus bounded by $2^b$.

	This covers all the loops in the algorithm.
	Since all of them are bounded, so is the algorithm as a whole, as it contains no blocking statements.
\end{proof}

\section{Belief Gain about Occurrences of Haps}
\label{sec:events}

In this section, we turn our attention to a sufficient condition for an agent
to establish belief that a group of reliable agents (a reliable agent will stay forever correct) has obtained belief about the correct occurrence of some event or action.
It follows already from \cref{thm:recv_phi} that sufficiently many disjoint hope chains
for $\varphi=\bigvee\limits_{\parbox{0.9cm}{\tiny\centering$G\subseteq\agents$,\\ $|G| = k$}}\bigwedge\limits_{j \in G} B_j \always{correct{j}} \wedge \trueoccurred{o}$, with $k + f \le n$, are enough for
establishing $B_i\varphi$. \cref{thm:mut_bel_occ} adds another condition, namely,
that among the disjoint hope chains for formula $\trueoccurred{o}$, at least $k$ are 
non faulty and hence truthfully deliver the information that some correct agent believes in $\trueoccurred{o}$.
Note that the two conditions are related but, in general, not identical.

\begin{theoremrep} \label{thm:mut_bel_occ}
For agent context $\chi \in \extension[B_f]$, interpreted system $\intsys = (\system{\chi}, \pi)$, run $r \in \system{\chi}$, timestamp $t \in \mathbb{N}$, action or event $o \in \haps$, agent $i \in \agents$, natural number $k \in \mathbb{N} \setminus \{0\}$ s.t. $k + f \le n$, and set $F \subseteq \agents$, which $i$ believes to be faulty, if
	\begin{gather} \label{eq:thm_bel_occ}
		\left(
		\begin{aligned}
			(\exists \Sigma' \in \disjss{\recvphi{\trueoccurred{o}}{r_i(t)}{i} \setminus F})\ |\Sigma'| \ &\ge \ k + f - |F| \quad \text{or } \\
			\biggl(\exists \Sigma'' \in \disjss{\recvphi{{\bigvee\limits_{\parbox{0.9cm}{\tiny\centering$G\subseteq\agents$,\\ $|G| = k$}}\bigwedge\limits_{j \in G} \always{\correct{j} \wedge B_j \trueoccurred{o}}}}{r_i(t)}{i} \setminus F}\biggr)\ |\Sigma''| \ &> \ f - |F|
		\end{aligned}
		\right)
		\\ \Rightarrow \ (\intsys, r, t) \models B_i \bigvee\limits_{\parbox{1.3cm}{\scriptsize\centering$G'\subseteq\agents$,\\ $|G'| = k$}}\bigwedge\limits_{j \in G'} \always{\correct{j} \wedge B_j \trueoccurred{o}}.
	\end{gather}
\end{theoremrep}
\begin{proof}
	The second line of the disjunction follows immediately from Theorem \ref{thm:recv_phi}.

	We prove the first line by contradiction.
	\begin{compactenum}
		\item $(\exists \Sigma' \in \disjss{\recvphi{\trueoccurred{o}}{r_i(t)}{i} \setminus F})\ |\Sigma'| \ \ge \ k + f - |F|$ \proofarg{by contr. assumption} \label{it:thm_mut_bel_occ_firstline_1}
	\item $(\intsys, r, t) \not\models B_i \bigvee\limits_{\parbox{1.3cm}{\scriptsize\centering$G'\subseteq\agents$,\\ $|G'| = k$}}\bigwedge\limits_{j \in G'} \always{\correct{j}} \wedge B_j \trueoccurred{o}$ \proofarg{by contr. assumption} \label{it:thm_mut_bel_occ_firstline_2}
	\item $\widetilde{r} \in \system{\chi} \text{ and } \widetilde{t} \in \mathbb{N} \text{ and } r(t) \pwrelation{i} \widetilde{r}(\widetilde{t}) \text{ and }(\intsys, \widetilde{r}, \widetilde{t}) \models \correct{i} \ \wedge \ \bigwedge\limits_{\parbox{1.3cm}{\scriptsize\centering$G'\subseteq\agents$,\\ $|G'| = k$}} \neg \bigwedge\limits_{j \in G'} \always{\correct{j}} \ \wedge$\\
		$B_j \trueoccurred{o}$ \proofarg{from line \eqref{it:thm_mut_bel_occ_firstline_2}, by sem. of $\not\models$, $\bigvee$, $\bigwedge$, $\neg$, $\neg B_i$, exist. inst.} \label{it:thm_mut_bel_occ_firstline_3}
		\item $|F| \le f$ \proofarg{from line \eqref{it:thm_mut_bel_occ_firstline_3}, by Lemma \ref{lem:corr_ag_bel_faulty_le_f}} \label{it:thm_mut_bel_occ_firstline_3a}
		\item $(\forall m \in F)\ (\intsys, r, t) \models B_i \faulty{m}$ \proofarg{by assumption} \label{it:thm_mut_bel_occ_firstline_4}
		\item $(\forall m \in F)\ (\intsys, \widetilde{r}, \widetilde{t}) \models \faulty{m}$ \proofarg{from lines \eqref{it:thm_mut_bel_occ_firstline_3}, \eqref{it:thm_mut_bel_occ_firstline_4}, by sem. of $B_i$ and univ. inst.} \label{it:thm_mut_bel_occ_firstline_5}
\item $\Sigma \in \disjss{\recvphi{\trueoccurred{o}}{\widetilde{r}_i(\widetilde{t})}{i} \setminus F} \text{ and } |\Sigma| \ \ge \ k + f - |F|$ \proofarg{from lines \eqref{it:thm_mut_bel_occ_firstline_1}, \eqref{it:thm_mut_bel_occ_firstline_3}, by sem. of $\pwrelation{i}$ and exist. inst.} \label{it:thm_mut_bel_occ_firstline_6}
		\item $f + k \le n$ \proofarg{by assumption} \label{it:thm_mut_bel_occ_firstline_6,5}
		\item $(\intsys, \widetilde{r}, \widetilde{t}) \models \bigvee\limits_{\parbox{1.3cm}{\scriptsize\centering$G'\subseteq\agents$,\\ $|G'| = k$}}\bigwedge\limits_{j \in G'} \always{\correct{j}}$ \proofarg{from line \eqref{it:thm_mut_bel_occ_firstline_6,5}, by at most $f$ faulty agents in runs of $\extension[B_f]$} \label{it:thm_mut_bel_occ_firstline_7}
		\item $\Sigma \setminus F = \Sigma \text{ and } \overline{\Sigma} \subseteq \Sigma \text{ and } |\overline{\Sigma}| \ge k \text{ and } (\forall \overline{\sigma} \in \overline{\Sigma})(\forall j \in \{1, \ldots, |\overline{\sigma}|\})\ (\intsys, \widetilde{r}, \widetilde{t}) \models \always{\correct{\pi_j\overline{\sigma}}}$ \proofarg{from lines \eqref{it:thm_mut_bel_occ_firstline_3}, \eqref{it:thm_mut_bel_occ_firstline_3a}, \eqref{it:thm_mut_bel_occ_firstline_5}, \eqref{it:thm_mut_bel_occ_firstline_6}, \eqref{it:thm_mut_bel_occ_firstline_7}, by at most $f$ faulty agents in runs of $\extension[B_f]$, sem. of $\le$, $\ge$, $\bigvee$, $\bigwedge$, exist. inst., using a pigeonhole argument} \label{it:thm_mut_bel_occ_firstline_8}
		\item $(\forall \overline{\sigma} \in \overline{\Sigma})(\exists \mu \in \msgsphi{\overline{H}_{\overline{\sigma}}\trueoccurred{o}}{\pi_1\overline{\sigma}}{i})\ (\intsys, \widetilde{r}, \widetilde{t}) \models \trueoccurred[i]{\recv{\pi_1\overline{\sigma}}{\mu}}$ \proofarg{from lines \eqref{it:thm_mut_bel_occ_firstline_6}, \eqref{it:thm_mut_bel_occ_firstline_8}, by sem. of $\wedge$, $\correct{i}$, $\trueoccurred[i]{\recv{\pi_1\overline{\sigma}}{\mu}}$ and Definition \ref{def:recv_phi} and \ref{def:seqSetDiff}} \label{it:thm_mut_bel_occ_firstline_9}
		\item $(\forall \overline{\sigma} \in \overline{\Sigma})(\exists \mu \in \msgsphi{\overline{H}_{\overline{\sigma}}\trueoccurred{o}}{\pi_1\overline{\sigma}}{i})\ (\intsys, \widetilde{r}, \widetilde{t}) \models \trueoccurred[\pi_1\overline{\sigma}]{\send{i}{\mu}}$ \proofarg{from lines \eqref{it:thm_mut_bel_occ_firstline_8}, \eqref{it:thm_mut_bel_occ_firstline_9}, by sem. of $\trueoccurred[\pi_1\overline{\sigma}]{\send{i}{\mu}}$, Definition of $\filtere[B_f]{}{}$} \label{it:thm_mut_bel_occ_firstline_10}
		\item $(\forall \overline{\sigma} \in \overline{\Sigma})\ (\intsys, \widetilde{r}, \widetilde{t}) \models B_{\pi_1\overline{\sigma}} \overline{H}_{\pi_2\overline{\sigma} \circ \ldots \circ \pi_{|\sigma|}\overline{\sigma}}\trueoccurred{o}$ \proofarg{from line \eqref{it:thm_mut_bel_occ_firstline_10}, by Definition \ref{def:phi_msgs}, Lemma \ref{lem:faulty_occ_pers}, Corollary \ref{cor:recv_phi_BH}, sem. of $\trueoccurred[j]{\send{i}{\mu}}$} \label{it:thm_mut_bel_occ_firstline_11}
		\item let $\widetilde{G} \ce \{j \in \agents \mid \overline{\sigma} \in \overline{\Sigma} \quad \text{and} \quad j = \pi_{|\overline{\sigma}|}\overline{\sigma}\}$ \proofarg{exist. inst.} \label{it:thm_mut_bel_occ_firstline_12}
		\item $|\widetilde{G}| \ge k$ \proofarg{from lines \eqref{it:thm_mut_bel_occ_firstline_6}, \eqref{it:thm_mut_bel_occ_firstline_8}, by sem. of $\subseteq$, Definition \ref{def:powSetOfMaxDisjSeqSets}} \label{it:thm_mut_bel_occ_firstline_13}
		\item let $\widetilde{G'} \subseteq \widetilde{G} \text{ and } |\widetilde{G'}| = k$ \proofarg{from line \eqref{it:thm_mut_bel_occ_firstline_13}, by sem. of $\subseteq$, exist. inst.} \label{it:thm_mut_bel_occ_firstline_13a}
		\item $(\forall j \in \widetilde{G'})\ (\intsys, \widetilde{r}, \widetilde{t}) \models \always{\correct{j}}$ \proofarg{from lines \eqref{it:thm_mut_bel_occ_firstline_8}, \eqref{it:thm_mut_bel_occ_firstline_12}, \eqref{it:thm_mut_bel_occ_firstline_13a}} \label{it:thm_mut_bel_occ_firstline_14}
		\item $(\forall j \in \widetilde{G'})\ (\intsys, \widetilde{r}, \widetilde{t}) \models B_j \trueoccurred{o}$ \proofarg{by lines \eqref{it:thm_mut_bel_occ_firstline_11}, \eqref{it:thm_mut_bel_occ_firstline_14}, by sem. of $H$, $\always{}$, $\correct{j}$, $\rightarrow$, reflexivity of $\pwrelation{j}$} \label{it:thm_mut_bel_occ_firstline_15}
		\item $(\intsys, \widetilde{r}, \widetilde{t}) \models \bigwedge\limits_{j \in \widetilde{G'}} \always{\correct{j}} \wedge B_j \trueoccurred{o}$ \proofarg{from line \eqref{it:thm_mut_bel_occ_firstline_11}, by sem. of $\forall$} \label{it:thm_mut_bel_occ_firstline_16}
		\item contradiction! \proofarg{from lines \eqref{it:thm_mut_bel_occ_firstline_3}, \eqref{it:thm_mut_bel_occ_firstline_13a}, by sem. of $\wedge$} \label{it:thm_mut_bel_occ_firstline_17}
	\end{compactenum}
\end{proof}

We conclude this section by noting that the first condition in \cref{thm:mut_bel_occ} could be strengthened by agent $i$ also considering a possible
occurrence of $\trueoccurred{o}$ in its own history, see
\cref{thm:loc_cond_bel_phi}, in which case $k$ can be
reduced by 1 if none of the hope chains in $\Sigma'$ contains $i$.

\section{Conclusions} \label{sec:concl}

We provided sufficient conditions for an agent to obtain belief of (1) the faultiness of (other)
agents and (2) of the occurrence of an event or action happening at some correct agent(s). Our
conditions work for any agent context where at most $f$ agents may be byzantine. They do not
require the agent to compute the underlying Kripke model, but can rather be checked locally
by the agent in bounded time just based on its current history.
Since protocols for
byzantine fault-tolerant distributed systems typically require an agent to detect (1)
and/or (2), our results are important stepping stones for the development of communication-efficient protocols and for proving them correct.

\paragraph{Acknowledgments.} We are grateful to Giorgio Cignarale, Hans van Ditmarsch, Stephan Felber, \\
Krisztina Fruzsa, Roman Kuznets, Rojo Randrianomentsoa and Hugo Rincon Galeana for many fruitful discussions.

\bibliographystyle{eptcs}
\bibliography{bib}

\end{document}

%% file: transition_function_WoLLIC.tex

\tikzset{nodes0/.style={anchor=west},}
\tikzset{nodes1/.style={text width=1.5cm,anchor=west},}
\tikzset{nodes2a/.style={text width=0.5cm,anchor=west},}
\tikzset{nodes2/.style={text width=0.8cm,anchor=west},}
\tikzset{nodes3/.style={text width=1.2cm,anchor=west},}
\tikzset{nodes4/.style={text width=1.3cm,anchor=west},}

\begin{tikzpicture}
    \node[nodes0] at (0,1) (0) {$\run{}{t}$};

    \node[nodes1] at (2.5,3)  (11) {$\agprotocol{n}{\run{n}{t}}$};
    \node[nodes1] at (2.5,2)  (01) {$\dots$};
    \node[nodes1] at (2.5,1)  (1) {$\agprotocol{1}{\run{1}{t}}$};
    \node[nodes1] at (2.5,-2) (2) {$\envprotocol{t}$};

    \draw[->,fill=white] (0) -- (11) node[midway,fill=white]{$\agprotocol{n}{}$};
    \draw[->] (0) -- (1) node[midway,fill=white]{$\agprotocol{1}{}$};
    \draw[->] (0) -- (2) node[midway,fill=white]{$\envprotocol{}$};

    \node[nodes2a] at (7,3)  (13a) {$X_n$};
    \node[nodes2a] at (7,2)  (03a) {$\dots$};
    \node[nodes2a] at (7,1)  (3a) {$X_1$};
    \node[nodes2a] at (7,-2) (4a) {$X_\epsilon$};
    
    \draw[->] (11) -- (13a) node[midway,fill=white] {$\adversary{}$};
    \draw[->] (1) -- (3a) node[midway,fill=white] {$\adversary{}$};
    \draw[->] (2) -- (4a) node[midway,fill=white] {$\adversary{}$};
    
    \node[nodes2] at (9.5,3)  (13) {$\alphaag{n}{r}{t}$};
    \node[nodes2] at (9.5,2)  (03) {$\dots$};
    \node[nodes2] at (9.5,1)  (3) {$\alphaag{1}{r}{t}$};
    \node[nodes2] at (12.5,3)  (13m) {$\alphaag{n}{r}{t}$};
    \node[nodes2] at (12.5,2)  (03m) {$\dots$};
    \node[nodes2] at (12.5,1)  (3m) {$\alphaag{1}{r}{t}$};

    \node[nodes2] at (9.5,-2) (4) {$X_\epsilon= \alphae{r}{t}$};
    
    \draw[->] (13a) -- (13) node[midway,fill=white] {$\mathit{global}$};
    \draw[->] (3a) -- (3) node[midway,fill=white] {$\mathit{global}$};
    \draw[double] (4a) -- (4) node[midway] {};
    \draw[double] (4) -- (4) node[midway] {};
    \node[nodes3] at (16,3) (15) {$\betaag{n}{r}{t}$};
    \node[nodes3] at (16,2) (05) {$\dots$};
    \node[nodes3] at (16,1) (5) {$\betaag{1}{r}{t}$};
    \node[nodes3] at (12.5,-2) (6m) {$\betae{r}{t}$};
    \node[nodes3] at (16,-2) (6) {$\betae{r}{t}$};
    
    \draw[double] (13) -- (13m);
    \draw[double] (3) -- (3m);
    \draw[double] (6) -- (6m);

    \draw[->] (13m) -- (15) node[midway,fill=white] (f3)  {$\filterag{n}{}{}$};
    \draw[->] (3m) -- (5) node[midway,fill=white] (f1) {$\filterag{1}{}{}$};
    \draw[->] (4) -- (6m) node[midway,fill=white] (fe) {$\filtere{}{}$};

    \draw[->] (13) edge (fe) (3) edge (fe);
    \draw[->,dotted] (13) edge (f1) (3) edge (f3);
    \draw[->,dashed] (6m) edge (f1) (6m) edge (f3);    
    
    \node[nodes4] at (19,3) (17) {$\run{n}{t+1}$};
    \node[nodes4] at (19,2) (07) {$\dots$};
    \node[nodes4] at (19,1) (7) {$\run{1}{t+1}$};
    \node[nodes4] at (19,-2) (8) {$\run{\epsilon}{t+1}$};

    \draw[->] (15) -- (17) node[midway,fill=white](19) {$\updateag{n}{}{}{}$};
    \draw[->,fill=white] (5) -- (7) node[midway,fill=white] (9){$\updateag{1}{}{}{}$};
    \draw[->] (6) -- (8) node[midway,fill=white] (a){$\updatee{}{}$};
    
    \draw[->] (5) -- (a);
    \draw[->] (15) -- (a);
    
    \draw[->,dashed] (6) -- (19) node[near end,fill=white] {$\betae[n]{r}{t}$};    
    \draw[->,dashed] (6) -- (9) node[near end,right,fill=white] {$\betae[1]{r}{t}$};

    \node[nodes0] at (21.5,1) (10){$\run{}{t+1}$};
    \draw[->] (7) edge (10) (17) edge (10) (8) edge (10);
    
    \draw[->] (-0.25,-3) -- (22.5,-3);
    \node at (0,-3)    (t0) {$|$};
    \node at (0,-3.5) {$t$};
    \node at (3.1,-3)    (t1) {$|$};
    \node at (7.3,-3)  (t2a) {$|$};
    \node at (9.8,-3)  (t2) {$|$};
    \node at (16.6,-3) (t3) {$|$};
    \node at (22,-3)   (t5) {$|$};
    \node at (22,-3.5) {$t+1$};

    \draw[-] (t0) -- (t1) node[midway,above] {Protocol phase};
    \draw[-] (t1) -- (t2a) node[midway,above] {Adversary phase};
    \draw[-] (t2a) -- (t2) node[midway,above] {Labeling phase};
    \draw[-] (t2) -- (t3) node[midway,above] {Filtering phase};
    \draw[-] (t3) -- (t5) node[midway,above] {Updating phase};
\end{tikzpicture}